\documentclass[prl,aps, article,amsfonts, nofootinbib, notitlepage]{revtex4-1}

\usepackage{caption}
\usepackage{subfig}
\usepackage{hyperref}
\usepackage[all]{hypcap}
\usepackage{epsfig}
\usepackage{amssymb}
\usepackage{amsthm}
\usepackage{amsmath}
\usepackage{amsmath}

\newcommand{\beq}{\begin{equation}}
\newcommand{\eeq}{\end{equation}}
\newcommand{\bea}{\begin{eqnarray}}
\newcommand{\eea}{\end{eqnarray}}

\newcommand{\nn}{\nonumber}
\newcommand{\benn}{\begin{displaymath}}
\newcommand{\eenn}{\end{displaymath}}

\newcommand{\tr}{{\rm Tr}}

\usepackage{multirow}
\usepackage{chemarrow}
\usepackage{amsmath}

\begin{document}
\begin{flushright}
{NT@UW-12-12}
\end{flushright}
\title{Charmed-Baryon Spectroscopy\\ from Lattice QCD with $N_{f}=2+1+1$ Flavors}

\author{Ra\'ul A. Brice\~no\footnote{{\tt briceno@uw.edu}}}
 \author{{Huey-Wen Lin}
 }
\affiliation{Department of Physics, University of Washington\\
Box 351560, Seattle, WA 98195, USA}

\author{Daniel R. Bolton
}
\affiliation{Department of Physics, Baylor University\\
One Bear Pl \#97316, Waco, TX 76798, USA}\

\begin{abstract}
We present the results of a calculation of the positive-parity ground-state charmed-baryon spectrum using $2+1+1$ flavors of dynamical quarks. The calculation uses a relativistic heavy-quark action for the valence charm quark, clover-Wilson fermions for the valence light and strange quarks, and HISQ sea quarks.
The spectrum is calculated with a lightest pion mass around $220$~MeV, and three lattice spacings ($a\approx0.12$~fm, $0.09$~fm, and $0.06$~fm) are used to extrapolate to the continuum. The light-quark mass extrapolation is performed using heavy-hadron chiral perturbation theory up to $\mathcal{O}(m_\pi^3)$ and at next-to-leading order in the heavy-quark mass.
For the well-measured charmed baryons, our results show consistency with the experimental values.
For the controversial $J=1/2$ $\Xi_{cc}$, we obtain the isospin-averaged value $M_{\Xi_{cc}}=3595(39)(20)(6)$~MeV (the three uncertainties are statistics, fitting-window systematic, and systematics from other lattice artifacts, such as lattice scale setting and pion-mass determination), which shows a $1.7~\sigma$ deviation from the experimental value.
We predict the yet-to-be-discovered doubly and triply charmed baryons $\Xi_{cc}^*$, $\Omega_{cc}$, $\Omega_{cc}^*$ and $\Omega_{ccc}$ to have masses $3648(42)(18)(7)$~MeV, $3679(40)(17)(5)$~MeV, $3765(43)(17)(5)$~MeV and $4761(52)(21)(6)$~MeV, respectively.
\end{abstract}

\maketitle

\section{Introduction}

In recent years, interest in charmed-baryon spectroscopy has resurfaced. This excitement has been partly triggered by the first observation of a candidate doubly charmed baryon $\Xi^+_{cc}(3520)$ by SELEX~\cite{SELEX1}, as well as its isospin partner $\Xi^{++}_{cc}(3460)$~\cite{SELEX2}. The SELEX Collaboration later confirmed their observation of $\Xi^+_{cc}(3520)$~\cite{SELEX_confirm}, but the BABAR~\cite{BABAR1}, BELLE~\cite{BELLE1}, and FOCUS~\cite{FOCUS} experiments have seen no evidence for either state of the isospin doublet $(\Xi^+_{cc},\Xi^{++}_{cc})$. The SELEX evidence for this doublet implies unprecedented dynamics, the most surprising of which is the $60$-MeV mass difference between the two states. All other previously observed isospin partners have mass differences one order of magnitude smaller.
The ground-state doubly charmed baryon has been previously studied theoretically via various methods, including: the nonrelativistic quark model~\cite{QM}, the relativistic three-quark model~\cite{RTQM}, the relativistic quark model~\cite{RQM}, QCD sum rules~\cite{QCDsum}, heavy-quark effective theory~\cite{HQET}, the Feynman-Hellmann theorem~\cite{FHT}, and lattice quantum chromodynamics (LQCD)~\cite{latt1, latt15, latt16,latt2, latt25,latt3,latt4}. Overall, theoretical predictions for this state suggest the $\Xi_{cc}$ mass to be $100$--$200$~MeV higher than that observed by SELEX.

There remain many undiscovered doubly and triply charmed baryon states. The recently upgraded Beijing Electron-Positron Collider (BEPCII) detector, the Beijing Spectrometer (BES-III), the LHC, and the future GSI project, the antiProton ANnihilation at DArmstadt (PANDA) experiment, will help further disentangle the heavy-baryon spectrum and resolve puzzles like the one mentioned above. LQCD calculations serve as direct first-principles theoretical input for these experiments.

Currently, LQCD provides the best option for performing reliable calculations of low-energy QCD observables. LQCD is a numerical calculation of QCD, which is necessarily performed in a finite discretized and Euclidean spacetime volume. These approximations introduce an infrared cut-off (the spatial extent $L$) and an ultraviolet cutoff (the lattice spacing $a$). The latter of these artifacts has been a source of large systematic errors in the heavy-quark sector of QCD. For heavy-quark masses satisfying $am_Q\ll{1}$, it is natural to control the discretization errors using nonrelativistic QCD (NRQCD)~\cite{NRQCD}. NRQCD has proven particularly useful when studying physics regarding the bottom quark, but for lattice spacings $\leq 0.12$~fm the charm-quark mass is too small to make the NRQCD approximation justifiable. Alternatively, one can implement relativistic heavy-quark actions~\cite{RHQ0, RHQ1, RHQ2, RHQ3,RHQ4}, where all $\mathcal{O}((am_Q )^n)$ corrections are systematically removed.

Several groups have performed lattice charmed-baryon calculations using the quenched approximation~\cite{latt3, latt4, latt5, latt6}. Although these have served as benchmark calculations of the charmed-baryon sector, the quenched approximation is a large source of systematic error that is difficult to estimate. Three previous groups have studied the charmed-baryon spectrum using dynamical quarks~\cite{latt1, latt15, latt16, latt2, latt25, latt8, Alexandrou:2012xk}.

Na~et~al.~\cite{latt2, latt25} performed a rather extensive calculation of charm- and bottom-baryon masses at three different lattice spacings ($a\approx0.15$~fm, $0.12$~fm, and $0.09$~fm). They used chiral perturbation theory ($\chi$PT)-inspired polynomial extrapolations of the light-quark masses but refrained from performing a continuum extrapolation of their results. From their results for the doubly charmed baryons, one could infer a $40$--$100$~MeV systematic error associated with discretization effects.

Liu~et~al.~\cite{latt1, latt15, latt16} did a rather nice exploratory calculation over four different pion masses and performed what is probably the best (to this day) chiral extrapolation of the $J={1}/{2}^+$ charmed-baryon spectrum using a relativistic heavy quark action for the charm quark. There are a few places where this calculation could be further improved. First, the lightest pion used in their calculation was about $290$~MeV; with advances in technology, we can get closer to the physical point. For baryons with no light degrees of freedom, this is a minor issue, but for isodoublet doubly charmed baryons the light-quark mass dependence is nontrivial. Second, they performed all calculations at a single coarse lattice spacing, $a\approx0.125$~fm, which lies near the upper limit of reliable spacings for studying charm physics. In their work, they used power-counting arguments to give estimates of the discretization effects. In particular, in the doubly charmed sector, they assigned a rather conservative systematic uncertainty associated with discretization effects, $\delta M_{h_{cc}}=-78$~MeV. This is by far their largest uncertainty across all states; for example, their result for the lightest doubly charmed baryon is $M_{\Xi_{cc}}= 3665(17)(14)^{+0}_{-78}$. Lastly, they restricted themselves to studying the $J={1}/{2}^+$ sector. The $J={1}/{2}^+$ and $J={3}/{2}^+$ charmed baryons are related by heavy-quark symmetries, which results in their chiral extrapolations being coupled. This is particularly relevant when performing a $\chi$PT-motivated extrapolation of the $(\Xi_{cc},\Xi_{cc}^*)$ doublet to the physical point.

The European Twisted-Mass (ETM) Collaboration recently presented results for $\Lambda_c$, $\Sigma_c$, $\Sigma^*_c$, $\Xi_{cc}$, $\Xi^*_{cc}$, and $\Omega_{ccc}$, 
using $N_f=2$ dynamical sea quarks with a lightest pion mass of about $260$~MeV and a relativistic action for the valence charm quark~\cite{Alexandrou:2012xk}. They used $\chi$PT-inspired polynomials for the light-quark mass extrapolation, neglecting $\mathcal{O}(1/m_Q)$ corrections and chiral-log contributions. Despite having performed calculations at three lattice spacings $a\in \{0.056(1),0.0666(6), 0.0885(6)\}$~fm, they refrained from extrapolating their results to the continuum. Instead, they estimated a discretization error that is incorporated into their systematics. Although historically, the use of $N_f=2$ dynamical sea quarks was a reasonable approximation, this (like full quenching) introduces a source of systematic error that can only be quantified when results are directly compared to $N_f=2+1$ or $N_f=2+1+1$ calculations.
 
In order to confidently deal with systematics due to discretization effects, it is necessary to perform calculations with highly improved actions, relativistic heavy-quark actions, and multiple lattice spacings in order to extrapolate to the continuum. With these goals in mind, we evaluated the positive-parity ground-state charm-baryon spectrum using two pion masses (with a lightest $m_\pi$ around 220~MeV) and three lattice spacings ($a\approx0.12$~fm, $0.09$~fm, and $0.06$~fm). In this work, we made three extensions to our previous preliminary calculation~\cite{Briceno}. Firstly, we used an ensemble at the super-fine $a\approx0.06$~fm lattice spacing in order to further constrain the continuum extrapolation. Secondly, when extrapolating the charmed-baryon masses to the physical $m_\pi$, we used heavy-hadron $\chi$PT (HH$\chi$PT)~\cite{hhchipt1, hhchipt2, hhchipt3, hhchipt4} at next-to-leading order (NLO) in $m_\pi$ and in the heavy-quark mass expansion, while in our previous work we had restricted ourselves to the LO $m_\pi$ dependence. In order to do this, we extended previous HH$\chi$PT results~\cite{hhchipt6, hhchipt7} to include $\mathcal{O}(1/m_Q)$ corrections. Thirdly, we quantified systematics associated with finite-volume effects, scale setting, the determination of $m_\pi$, $\mathcal{O}(m^4_\pi,a^2m_\pi)$ corrections to the expressions used to extrapolate to the physical point, and the strange-mass tuning.

This paper is structured as follows. In Sec.~\ref{formulation}, we outline the formulation of the lattice calculation, including the actions used for the sea, valence light, and valence charm quarks, as well our procedure for setting the scale independently, and the construction of our correlation functions. In Sec.~\ref{ccbar}, we present the tuning of the charm-quark action and show the results for the charmonium spectrum. In this section, we present the results for the $m_{D_s}-m_{\eta_c/2}$ splitting, which is shown to have rather large lattice-spacing dependence, but the result presented is in agreement with experiment when extrapolated to the continuum. Section~\ref{cbs} outlines our analysis of the charmed-baryon spectrum and includes a detailed discussion of the $\mathcal{O}(m^3_\pi, 1/m_Q)$ HH$\chi$PT expressions for the masses. In this section, the $a$ dependence of the charmed-baryon sector is discussed, as well the systematics mentioned at the end of the previous paragraph. Finally, in Sec.~\ref{conclusion} we give a summary of our results and a comparison of the yet-to-be-discovered masses across different models.

\section{Lattice Formulation \label{formulation}}
 \subsection{Light-Quark Action \label{lqaction}}

In this work, we used $N_{f}=2+1+1$ gauge configurations that were generated by the MILC Collaboration with the highly improved staggered quark (HISQ)~\cite{MILC, MILC2,HPQCD, HPQCDUKQCD, meson4} action for the sea quarks. The implementation of the HISQ action, first proposed by the HPQCD/UKQCD Collaboration~\cite{HPQCD, HPQCDUKQCD, meson4}, has been shown to further reduce lattice artifacts as compared to the asqtad action~\cite{MILC}. Staggered actions reduce the number of doublers to four ``tastes'', which are reduced to the desired number of true flavors by taking the fourth-root of the fermionic determinant. As a result, staggered actions have two sources of discretization errors. The first is due to the discretization of the derivative, while the second is associated with taste-exchange interactions in quark-quark scattering. It has been shown that the latter type of errors are suppressed at $\ll1\%$ level when the HISQ action is used for light quarks at lattice spacings of 0.1~fm or less~\cite{meson4}. Furthermore, its suppression of $\mathcal{O}((am)^4)$ errors makes the HISQ action a desirable candidate for studying charm physics on the lattice~\cite{meson4}. 
Lastly, despite the HISQ action being significantly more computationally expensive than the asqtad action~\cite{hisqvasqtad}, it is still more economical than a non-staggered action. This has allowed the MILC Collaboration to recently generate multiple $N_{f}=2+1+1$ HISQ ensembles, with a range of lattice spacings $a\in [0.045,0.15]$~fm and three light-quark (up, down) masses corresponding to $m_\pi\in \{140, 220, 310\}$~MeV. This variety of ensembles allows for clean extrapolations to the physical pion mass and the continuum limit.

Hypercubic blocking~\cite{HYP} is implemented on the gauge configurations
in order to further reduce the ultraviolet noise from the gauge field.
For the valence light (up, down and strange) fermions a tree-level tadpole-improved clover-Wilson action is used \footnote{The light
clover propagators were generated and provided by the PNDME Collaboration \cite{Lin:2011zz, Gupta:2012rf, Bhattacharya:2012wk}.}, since the
construction of baryon operators with staggered fermions is rather complicated.
However, for the coarser and lighter pion mass ensembles (such as 140-MeV pion mass at $0.12$~fm), one runs into the problem of exceptional configurations where the clover-Dirac operator has near-zero modes~\cite{exceptional}. Thus, in this work, we were limited to heavier light-quark masses which correspond to $m_\pi\in \{220, 310\}$~MeV with lattice spacings of around 0.06, 0.09 and 0.12~fm.

Because the actions used for the sea and valence quarks differ, the calculation presented here is partially quenched and violates unitarity. In order to restore unitarity, it is necessary to match the valence- and sea-quark masses, as well as to extrapolate the results to the continuum. Due to the four-fold degeneracy of the staggered action, in the continuum limit it has an $SU(8)_L\otimes SU(8)_R\otimes U(1)_V$ chiral symmetry. In this limit, each pion obtains 15 degenerate partners. A finite lattice spacing breaks this symmetry and lifts the degeneracy~\cite{Orginos:1999cr, *Orginos:1998ue, *Toussaint:1998sa, *Lagae:1998pe, *Lepage:1998vj}. Therefore, there is an ambiguity when tuning the valence-quark mass to the sea-quark mass.
We chose to simultaneously tune the light- and strange-quark masses to assure that the valence pion and kaon masses match those of the lightest Goldstone Kogut-Susskind sea pion and kaon masses, as shown in Table~\ref{ensembles}. The Goldstone Kogut-Susskind sea pion is the lightest pion, the only one that becomes massless in the chiral limit for a nonzero lattice spacing. Ideally, one would want to perform all calculations at a range of light, strange and charm masses and simultaneously extrapolate all masses to their physical values. Due to limited in computational resources, we performed calculations at a single strange quark mass, but as will be discussed in Sec.~\ref{spacing} our determination of $m_K$ at the continuum and physical $m_\pi$ is in agreement with experiment. This gives us confidence that the strange-quark mass is tuned properly.

\begin{center}
\begin{table}
\label{tab:param1}
\begin{tabular}{|c|ccccccccc|}
\hline

& \hspace{.1cm}$\beta$\hspace{.1cm}& \hspace{.1cm}$(am_\pi)_{\rm{sea}}$\hspace{.1cm}& \hspace{.1cm}$(am_K)_{\rm{sea}}$\hspace{.1cm}&
\hspace{.1cm}$(am_\pi)_{\rm{val}}$\hspace{.1cm}& \hspace{.1cm}$(am_K)_{\rm{val}}$\hspace{.1cm}&
\hspace{.1cm}${L}^3\times {T}$\hspace{.1cm}&\hspace{.1cm}${L}m_\pi$\hspace{.1cm}
&\hspace{.1cm}$N_\text{cfgs}$\hspace{.1cm}&\hspace{.1cm}$N_\text{props}$\hspace{.1cm} \\\hline \hline
\multirow{1}{*}{\hspace{.2cm}\textbf{A1}\hspace{.2cm}} &$
6.00$&0.18931(10)&0.32375(12) &0.18850(79)(55)&0.32358(58)(67)&$24^3\times 64$&$4.5$&504&2016 \\\hline
\multirow{1}{*}{\textbf{A2}} &$6.00$&0.13407(6)&0.30806(9)&0.13584(79)(59)&0.30894(52)(60)&$32^3\times 64$&$4.4$&477&1908  \\\hline
\multirow{1}{*}{\textbf{B1}} &$6.30$& 0.14066(13)&0.24085(14)
&0.14050(40)(28)&0.24032(39)(23)&$32^3\times 96$&$4.5$&391&1564\\\hline
 \multirow{1}{*}{\textbf{B2}} &$6.30$& 0.09845(9)& 0.22670(12)&0.09950(53)(23)&0.22464(27)(35)&$48^3\times 96 $&$4.8$&432&1568 \\\hline
 \multirow{1}{*}{\textbf{C1}} &6.72& 0.09444(9)&0.16204(11)&0.09444(38)(9)&0.16086(29)(68)&$48^3\times 144 $&$4.5$&330&1320
\\\hline\hline
\end{tabular}
\caption{Details of the configurations and propagators used in this work. The subscript ``sea'' labels the lightest sea pseudoscalar masses from the HISQ action~\cite{MILC, MILC2}, while the subscript ``val'' labels the valence masses. The sea hadron masses have a single uncertainty due to the statistics, while the valence masses include statistical and systematic uncertainty due to fitting-window selection as defined in Sec.~\ref{fitting}. Additionally, listed are the spatial ($L$) and temporal extents ($T$) in lattice units, the value of $m_\pi L$, the number of configurations, and the number of measurements performed for each ensemble. }
\label{ensembles}
\end{table}
\end{center}

\subsection{Correlation Functions and Fitting Method \label{fitting}}
Before discussing the tuning of the charm-quark action, let us explain how we constructed our correlation functions and extracted hadronic masses.
For a given interpolating hadron operator, ${O}^{(i)}_{H}$, we construct the two-point correlation functions
\begin{eqnarray}
\label{correlation}
C^{(ij)}_{H}(t-t_0)=\sum_{\textbf x}\langle\mathcal{O}^{(i)}_{H}(t,\textbf x)\mathcal{O}^{(j)\dag}_{H}(t_0,\textbf x_0)\rangle,
\end{eqnarray}
where the superscripts $i$ and $j$ label the smearing type of the annihilation and creation operator, respectively, $\{\textbf x_0, t_0\}$ labels source location, and $\{\textbf x, t\}$ the sink location. In order to reduce statistical noise, the two-point functions are averaged over four source locations for each gauge configuration.

Both the baryonic and mesonic correlation functions are calculated with gauge-invariant Gaussian-smeared (S) sources and point (P) sinks.
For the mesons, we use the generalized Prony-Matrix (PM) method~\cite{prony} over the smeared-smeared (SS) and smeared-point (SP) correlation functions. The PM method uses the fact that each choice of smearing parameters corresponds to a particular linear combination of the exponentiated masses ($m_j$) and the corresponding overlap factors ($A_j$), $C^{(i)}_H(t)={A}_0^{(i)}e^{-m_0t}+{A}_1^{(i)}e^{-m_1t}+\cdots$. By computing correlation functions with two sets of smearing parameters, we can determine the two lowest energy states that have overlap with the interpolating operator used by solving the eigenvalue equation
\begin{eqnarray}
My_H(t+1)-Vy_H(t)=0
\end{eqnarray}
where $y^T_H(t)=(C^{\rm(SS)}_H(t),C^{\rm (SP)}_H(t))$. One solution to this equation is given by~\cite{prony}
\begin{eqnarray}
M=\left[\sum^{\tau+t_W}_{t=\tau}y_H(t+1)y^T_H(t)\right]^{-1}
\hspace{1cm}
V=\left[\sum^{\tau+t_W}_{t=\tau}y_H(t)y^T_H(t)\right]^{-1},
\end{eqnarray}
where the window size $t_W$ must be $\geq1$ in order to ensure the matrices within the brackets are invertible. For each hadron, $\tau$ is chosen in order to maximize the plateau of the ground state. The statistical uncertainties of the extracted hadron masses are evaluated using the jackknife method.

We test the PM method for a subset of the baryonic masses and compare the results with those extracted from a single-exponential and double-exponential fits to the SP correlation function at large Euclidean time. We find these to be in agreement within our systematics, with the single-exponential having the smaller uncertainty. As a result, we choose to extract all masses from the single-exponential behavior of the SP correlation function.

For all energies extracted, we determine the statistical uncertainty and a systematic associated with choosing a fitting window $[t_i,t_f]$. In order to estimate the latter, for all fitting windows that fall within $[t_i,t_f+2]$ we calculate the energy, $\chi^2$, and goodness of fit ${Q(d)}$ (defined as ${(2^{d/2}\Gamma(d/2))^{-1}}\int^{\infty}_{\chi^2}d\chi_0^2(\chi_0^2)^{d/2-1}e^{-\chi_0^2/2}$), which depends on the number of degrees of freedom $d$ and is optimally near 1. From this ensemble of energies, we define the systematic as the standard deviation of the energies weighted by $Q(d)$.

 \subsection{Charm-Quark Action \label{hqaction}}

Since the charm-quark mass is too light to justifiably implement a nonrelativistic action for the lattice spacings used in our calculation, it is necessary to use a relativistic action. To systematically remove the $\mathcal{O}((m_{c}{a})^n)$ discretization artifacts (where $m_{c}$ is the charm-quark mass), we use the following relativistic heavy-quark action for the valence charm quark~\cite{RHQ1, RHQ2, RHQ3,RHQ4}:
 \begin{eqnarray}
S_{{Q}}&=&\sum_{{x,x'}}\overline{{Q}}_x\left(m_0+\gamma_0D_0-\frac{a}{2}D_0^2+\nu\left(\gamma_iD_i-\frac{a}{2}D_i^2\right)
-\frac{a}{4}c_\text{B}\sigma_{{ij}}{F}_{{ij}}
-\frac{a}{2}c_\text{E}\sigma_{{0i}}{F}_{{0i}}\right)_{xx'}{Q}_{x'},
\end{eqnarray}
where ${Q}_x$ is the heavy-quark field at the site $x$, $\gamma_{\nu}$ are the Hermitian gamma matrices that satisfy the Euclidean Clifford algebra $\sigma_{\mu\nu}={i}[\gamma_{\mu},\gamma_{\nu}]/{2}$, $D_\mu$ is the first-order lattice derivative, and $F_{\mu\nu}$ is the Yang-Mills field-strength tensor. The parameters $\{m_0, \nu, c_\text{B}, c_\text{E}\}$ must be tuned to assure $\mathcal{O}((m_{c}{a})^n)$ terms have been removed. For the coefficients $c_\text{B}$ and $c_\text{E}$ we use the tree-level tadpole-improved results~\cite{latt1, latt15, latt16, clover_terms} $c_\text{B}={\nu}/{u_0^3}, c_\text{E}={1+\nu}/{(2u_0^3)}$ with the tadpole factor $u_0$ defined as $u_0=({1}/{3})\left\langle\sum_{p} \tr{\left({U}_{{p}}\right)}\right\rangle^{1/4}$, where $U_p$ is the product of gauge links around the fundamental lattice plaquette $p$.

\begin{figure}
\begin{center}
\includegraphics[totalheight=6cm]{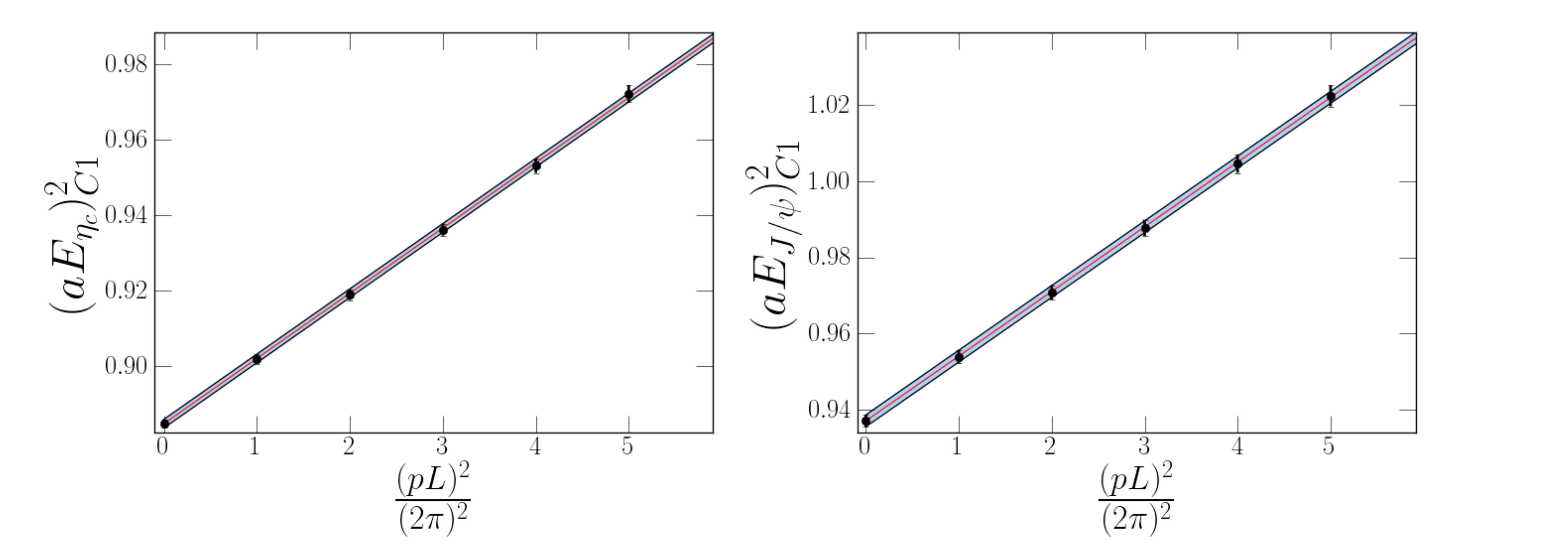}
\end{center}
\caption{A sample dispersion relation for the $\eta_c$ and $J/\psi$. The six points correspond to energies (and uncertainties) for the at the six lowest-momenta states: $(0,0,0),$ $(1,0,0),$ $(1,1,0)$, $(1,1,1)$, $(2,0,0)$, $(2,1,0)$ in units of $({2\pi}/{L})a^{-1}$, and their permutations. The red line is the resulting fit to the data using the relativistic dispersion relation ${E}_{{H}}^2=m_{{H}}^2+ c^2{p}^2$, and the blue band includes the statistical and systematic errors added in quadrature. The energies shown are obtained using the full statistics of the $\textbf{C1}$ ensemble and have been extrapolated to the physical charm mass. From the fit we obtain the speed of light and its statistical and systematic uncertainties, $c_{\eta_c}=1.0039(28)(9)$ and $c_{J/\psi}=0.9964(35)(5)$.}\label{dispersion}
\end{figure}

The coefficients $m_0$ and $\nu$ were simultaneously determined nonperturbatively by requiring the ratio ${m_{\overline{1S}}}/{m_{\Omega}}\equiv({m_{\eta_c}+3m_{J/\psi}})/{(4m_{\Omega})}$ to be equal to its experimental value, 1.83429(56), and \{$\eta_c$, $J/\psi$\} to satisfy the correct dispersion relation, ${E}_{{H}}^2=m_{{H}}^2+ {p}^2$. In constructing the charmonium correlation functions, we used the local interpolating operators shown in Table~\ref{cc_oper}.
The dispersion relation was matched using $\eta_c$ and $J/\psi$ energies at the six lowest momenta: $(0,0,0),$ $(1,0,0),$ $(1,1,0)$, $(1,1,1)$, $(2,0,0)$, $(2,1,0)$ in units of $({2\pi}/{L})a^{-1}$, and their rotational equivalents. In practice, we performed the initial tuning with a subset of 40 gauge configurations (with four sources each).
Clearly this procedure does not guarantee correct tuning upon analysis of the full ensemble. Therefore, we used two separate charm-quark masses and extrapolated to the physical charmonium mass. These two points allowed us to interpolate linearly in $am_{c}$ to the physical charm-quark mass defined by ${m_{\overline{1S}}}/{m_{\Omega}}= 1.83429(56)$. The valence charm-quark masses used for each ensemble are shown in Table~\ref{ensembles2}. Figure ~\ref{dispersion} shows examples of the resulting dispersion relations for the ${\eta_c}$ and ${J/\psi}$ with full statistics after extrapolating to the physical charm mass from one of the ensembles, $\textbf{C1}$, and they show that the slopes are consistent with 1.

 \begin{center}
\begin{table}
\label{tab:param2}
\begin{tabular}{|c|cccc|}
\hline

& \hspace{.1cm}$a~[\text{fm}]$\hspace{.1cm}& \hspace{.1cm}$am_\Omega$\hspace{.1cm}&\hspace{.1cm}$am_{c1}$\hspace{.1cm}&\hspace{.1cm}$am_{c2}$\hspace{.1cm} \\\hline \hline
\multirow{1}{*}{\hspace{.2cm}\textbf{A1}\hspace{.2cm}} &0.11926(77)(51)&
1.0291(56)(37)
&0.901&0.872 \\\hline
\multirow{1}{*}{\textbf{A2}} &$0.11926(77)(51)$
&1.0192(31)(21)
&0.900& 0.853  \\\hline
\multirow{1}{*}{\textbf{B1}} &$0.0871(10)(5)$
&0.7562(81)(52)
&0.561&0.536\\\hline
 \multirow{1}{*}{\textbf{B2}} &$0.0871(10)(5)$&
 0.7463(52)(25)
 &0.552& 0.522\\\hline
 \multirow{1}{*}{\textbf{C1}} & $0.0578(13)(19)
$&0.5148(17)(39)&0.319& 0.309

\\\hline\hline
\end{tabular}
\caption{The lattice spacings and $\Omega$ masses cited include the statistical and systematic uncertainties due to the fitting window.
The lattice spacings are determined by the chiral extrapolation of the $\Omega$ mass to the physical value of $(m_\pi/m_\Omega)^2$ for each value of $\beta$\footnote{Note that we calculate the $\Omega$ mass $[am_\Omega=0.5007(65)(96)]$ on 200 configurations for $a\approx 0.06$~fm, 220-MeV to fix the lattice spacing for ensemble $\textbf{C1}$.}.
 Additionally listed are the two bare masses of the valence charm quarks used for each ensemble.}
\label{ensembles2}
\end{table}
\end{center}

 \subsection{Lattice-Spacing Determination and Discussion of $m_H/m_\Omega$ Ratios \label{spacing}}
 \begin{figure}
\begin{center}
\includegraphics[totalheight=5cm]{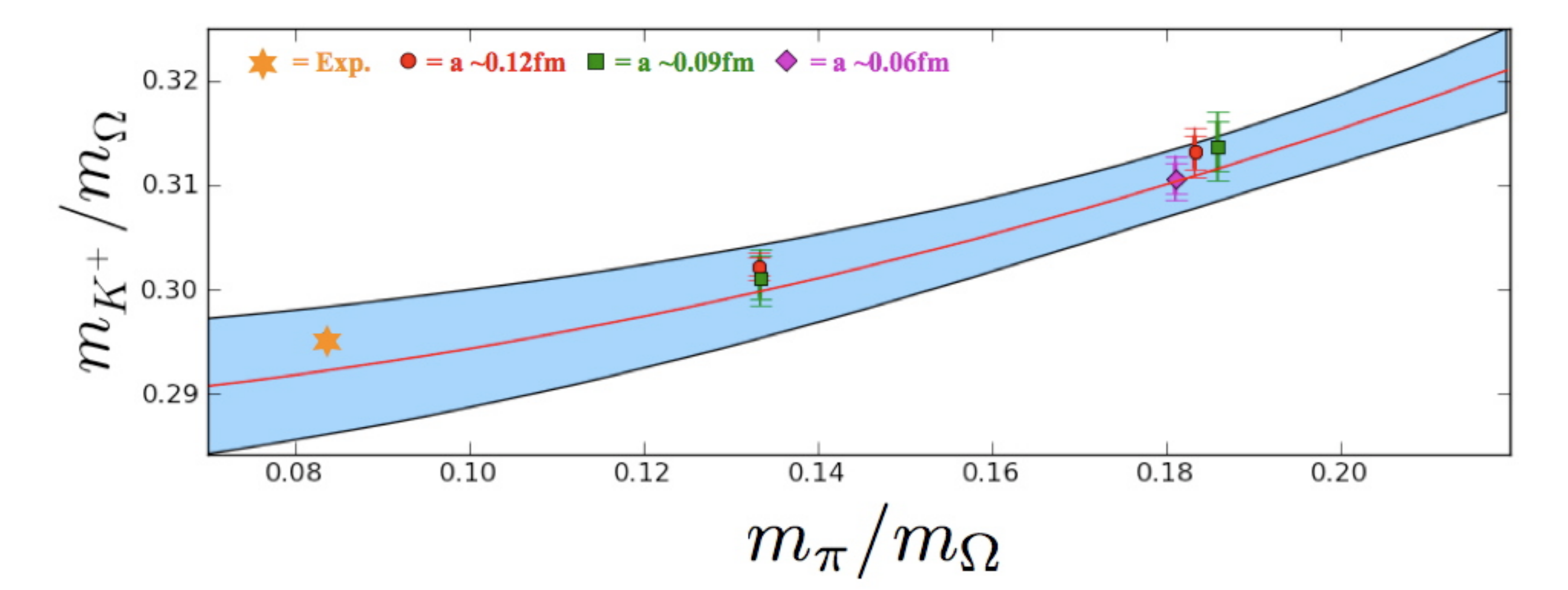}
\end{center}
\caption{$\chi$PT and continuum extrapolations of the kaon mass. The red line indicates the fit of the data that has been extrapolated to $a=0$. The blue band includes the statistical and systematic errors added in quadrature.}\label{kaon_plot}
\end{figure}
As mentioned earlier, it is necessary to evaluate the spectrum at multiple lattice spacings in order to simultaneously restore unitarity and remove discretization errors. With this in mind, we
perform the calculation at three lattice spacings, $a\approx 0.06$~fm, 0.09~fm and 0.12~fm. For the coarse ($a\approx0.12$~fm) and fine ($a\approx0.09$~fm) lattice spacings, we use two different light-quark masses corresponding to $m_\pi\approx 220, 310$~MeV; for the super-fine ($a\approx0.06$~fm) ensemble we use a single light quark, $m_\pi\approx310$~MeV. We calculate the $\Omega$ mass on 200 configurations for $a\approx 0.06$~fm and $m_\pi\approx220$~MeV to fix the lattice spacing for ensemble $\textbf{C1}$.

In order to obtain physical masses in the continuum, it is necessary to determine the lattice spacing for the five ensembles used. Currently, the most precise determination of lattice spacings for the MILC ensembles is by the HPQCD Collaboration~\cite{HPQCD}; however, their determinations of the lattice spacings for the \textbf{B2} and \textbf{C1} ensembles remain unpublished. For this reason, we perform our own determination. Due to the small ${m_{\pi}^2}$-dependence of $m_\Omega$ (at the few-percent level) we choose to set the scale by extrapolating the value of $am_\Omega$ across all ensembles with the same value of $\beta$ to the physical pion mass. We define the lattice spacing by dividing  $am^{\rm{phys}}_\Omega$ by the physical $\Omega$ mass, $1672.45(49)$~MeV.

In constructing the correlation functions for the $\Omega$, we use $(\Omega)^{{i}}=
\epsilon^{{klm}}{P}^+({P}^{3/2}_{{E}})^{{ij}}{q}^{{k}}_{{s}}\left({q}^{{lT}}_{{s}}\Gamma^{{j}}{q}^{m}_{{s}}\right)$ as the interpolating operator. The strange-quark annihilation operator is denoted ${q}^k_{s}$ with color index $k$, $\Gamma^{{i}}=C\gamma^i$ are the symmetric spin matrices (where $C$ is the charge-conjugation matrix), ${P}^+={(1+\gamma^4)}/{2}$ is the positive-parity projection operator, and $({P}^{3/2}_{E})^{ij}= \delta^{ij}-\frac{1}{3}\gamma^{i}\gamma^{j}$ are the spin-projection operators for spin-$3/2$ particles.

One can determine $m_\Omega$ as a function of ${m_{\pi}^2}$ via $SU(3)$ $\chi$PT, but this expression suffers from rather large expansion parameters ($m_K/\Lambda_\chi$, $m_\eta/\Lambda_\chi$) and does not always describe lattice baryon masses well. Alternatively, it has been proposed that the hyperon masses can be extrapolated using a two-flavor chiral perturbation theory~\cite{su2}. With a faster convergence than its three-flavor counterpart, the advantages of this approach are clear. The cost is manifested in a larger set of unknown coefficients. Using $SU(2)$ $\chi$PT for the hyperons, the $\Omega$ mass as a function of ${m_{\pi}^2}$ up to $\mathcal{O}(m_\pi^6)$ is~\cite{su2}
\begin{eqnarray}
\label{omega}
m_{\Omega}=m_{\Omega}^0+\frac{m_\pi^2}{4\pi f_\pi}\sigma^{(2)}_\Omega
+\frac{m_\pi^4}{(4\pi f_\pi)^3}\left[\sigma^{(4)}_\Omega\log{\frac{m^2_\pi}{\mu^2}}+\beta^{(4)}_\Omega\right]
+\frac{m_\pi^6}{(4\pi f_\pi)^5}\left[\sigma^{(6)}_\Omega\log^2{\frac{m^2_\pi}{\mu^2}}+\beta^{(6)}_\Omega+\gamma_{\Omega}^{(6)}\right],
\end{eqnarray}
where $f_\pi=130.7(4)$~MeV is the pion decay constant, and the \{$\sigma_\Omega,\beta_\Omega, \gamma_\Omega$\} are the low-energy coefficients (LECs) of the theory. Because at each lattice spacing we have (at most) two ensembles with two corresponding values of $m_\pi$, we are forced to truncate Eq.~\ref{omega} at $\mathcal{O}(m_\pi^2)$ in order to retain a reasonable level of precision. This truncation introduces a systematic uncertainty into our calculations which will be accounted for in Sec.~\ref{system}.

Further details of the ensembles, including our determination of the lattice spacing and the $\Omega$ mass are listed in Table~\ref{ensembles2}. The values determined by the MILC Collaboration are $a=0.1211(2)$~fm for the coarse and $a=0.0884(2)$~fm for the fine. The HPQCD Collaboration performed a rather extensive program in which they determined the lattice spacing for each ensemble using three different quantities: Upsilon $2S$-$1S$ splitting, the decay constant of the $\eta_s$ meson, and the $r_1/a$ ratio~\cite{HPQCD}. We determine a single lattice spacing for each value of $\beta$ and find central values that are consistently below both the MILC and HPQCD central values. This difference in the definition of the lattice spacing should have no impact on continuum-extrapolated results.

Table~\ref{ensembles2} shows that the lattice spacing for the ensemble $\textbf{C1}$ is currently determined at the $\sim  4\%$ level of precision. For the same reasons discussed above, we choose to determine the physical hadron masses using the $m_H/m_{\Omega}$ ratio. As will be shown, the $m_H/m_{\Omega}$ is determined at the sub-$1\%$ level for all ensembles and particles. Due to the removal of the $\mathcal{O}(m_\pi^4)$-terms in Eq.~\ref{omega}, we proceed to truncate all of our chiral fits at the $\mathcal{O}(m_\pi^3)$ level of accuracy, and estimate a systematic error associated with this approximation (see Sec.~\ref{system}).

Because we are using the strange mass to set the scale, it is important to first test the strange-mass tuning, which we do using the kaon mass. For all the pseudoscalar mesons, we use the standard local operators $\mathcal{O}_{H}={\bar q}^k_{f}\gamma_5q^k_{f'}$, where $q^k_{f}$ is the annihilation operator for a quark with flavor $f$ and color index $k$. As discussed in Ref.~\cite{kchipt}, when reducing the symmetry of $\chi$PT from $SU(3)$ to $SU(2)$, kaons can be represented as a matter field that couples to the $SU(2)$ chiral currents. This treatment of the kaons is referred to as $K\chi$PT. The advantage of $K\chi$PT is that the largest expansion parameter is $m_\pi^2/m_K^2<m_K^2/(4\pi f_\pi)^2$.
Using $K\chi$PT, the kaon mass as a function of $m_\pi$ is found to be~\cite{kchipt}
\begin{eqnarray}
\label{lambda}
\frac{m_K}{m_{\Omega}}&=&\frac{m^0_K}{m^0_{\Omega}}-\left({\sigma_K}+\frac{m_K{\sigma_{\Omega}}}{4\pi f_\pi m_\Omega}\right)\frac{m^2_{\pi}}{m_{\Omega}}+\mathcal{O}(m^4_{\pi}),
\end{eqnarray}
where $m^0_K$ is the bare kaon mass, and the $a$-dependence is parametrized by $c_a (m^{\rm{phys}}_\Omega a)^2$. For the kaon and for all other hadrons studied in this work, the continuum-limit mass is recovered by multiplying the ratio at the physical point by the physical $\Omega$ mass, $1672.45(49)$~MeV.

In Table~\ref{ensembles} the valence kaon masses are shown for each ensemble. In Fig.~\ref{kaon_plot} we show the values for the kaon mass for each ensemble with the corresponding statistical and systematic uncertainties as a function of $m_\pi/m_\Omega$, as well as the chiral extrapolation at the continuum. Figure~\ref{kaon_plot} shows that the lattice-spacing dependence of the kaon is rather small, and that the extrapolated value, $m_{K^+}=488.7(5.3)(5.3)(5.7)$~MeV (the three uncertainties are statistics, fitting-window systematic, and systematics from scale setting, $\mathcal{O}(m^4_\pi,a^2m_\pi)$-corrections to the expressions used to extrapolate to the physical point, finite volume, and strange-mass tuning as discussed in Sec.~\ref{system}), agrees with experiment within our systematics. This confirms our strange-mass tuning as well as our scale determination and extrapolation procedure using the $m_H/m_\Omega$ ratio.


\section{Charmonium Spectrum\label{ccbar}}
\begin{figure}
\begin{center}
\includegraphics[totalheight=2cm]{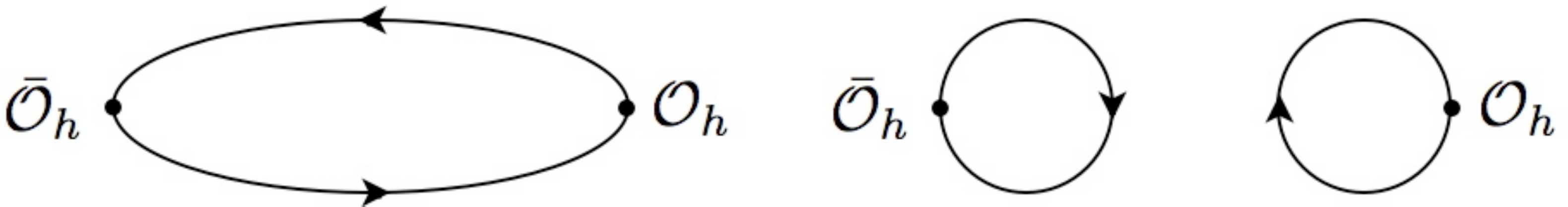}
\end{center}
\caption{Diagrams that contribute to the charmonium correlation functions. In this work we evaluate the contribution from connected diagrams (left) and neglect disconnected diagrams (right). The latter are OZI suppressed, and previous lattice calculations have determined their contributions to be consistent with zero~\cite{disc, disc2, disc2, disc3}. }
\label{diag}
\end{figure}

In this section, we calculate the charmonium $1S$ splitting and the rest of the charmonium spectrum in the continuum limit, and we compare them with experimental and previous dynamical lattice results.
We use the ratios of spin averages of $\eta_c$ and $J/\psi$ masses to $\Omega$ baryon masses to tune the charm-quark mass for each ensemble; thus, the splitting between them is not fixed in our calculations. Any deviations from the well-measured experimental values give us an estimation of the final systematics. 

\begin{figure}
\begin{center}
\includegraphics[totalheight=8.5cm]{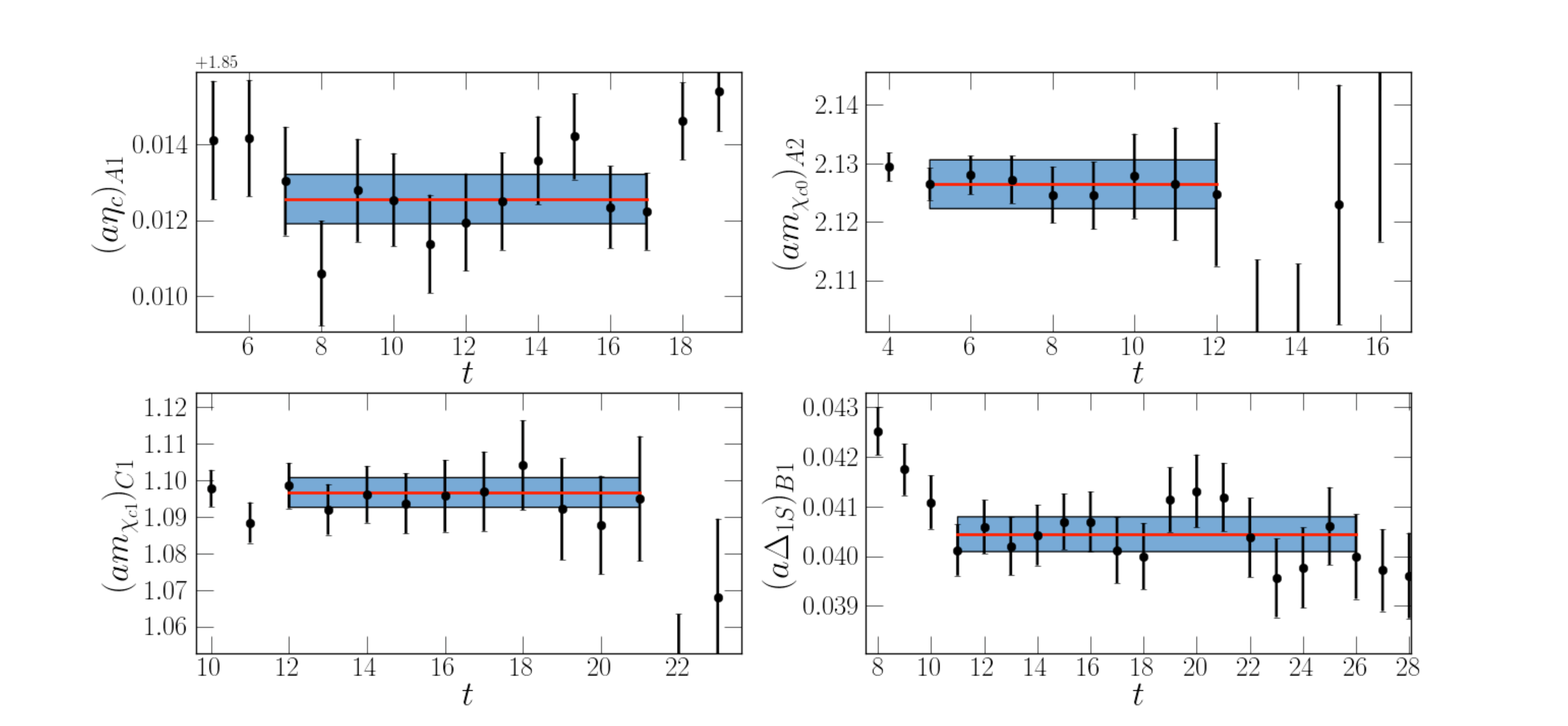}
\end{center}
\caption{Sample effective-mass plots of the charmonium spectrum from the various ensembles. The errorbar shown includes the statistical and systematic uncertainty (from varying the fitted range) added in quadrature.  }\label{cc_emps}
\end{figure}

In constructing the meson correlation functions, we restrict ourselves to the local interpolating operators shown in Table~\ref{cc_oper}. In order to evaluate the full correlation functions of the charmonium spectrum, we need to perform two different types of propagator contractions, as depicted in Fig.~\ref{diag}, connected and disconnected diagrams. Disconnected diagrams would increase the number of propagators needed by approximately two orders of magnitude but are suppressed by the OZI rule~\cite{OZI1, *OZI2, *OZI3, *OZI4}. Previous lattice calculations at zero temperature have shown disconnected diagrams in the charmed sector are rather noisy, and their contributions to the hyperfine splitting are in the range of 1--4~MeV and consistent with zero~\cite{disc, disc2, disc2, disc3}. Thus, we neglect contributions arising from disconnected diagrams here. Figure~\ref{cc_emps} displays examples of the effective-mass plots after performing the generalized Prony-Matrix method for the charmonium sector, and the charmonium masses for each ensemble are shown in Table~\ref{cc_ens} in lattice units.

For every hadron, we calculated the ratio of its mass to the $\Omega$ mass, $m_{{H}}/{m_{\Omega}}$ at the two different values of the charm-quark mass. After interpolating these to the physical charm-quark mass for each ensemble, we simultaneously extrapolated the five values of the hadron masses to the continuum and the physical $m_\pi$. To perform the light-quark mass extrapolation we use the $SU(2)$ $\chi$PT expressions, which up to $\mathcal{O}(m_\pi^3)$ is linear in $m_\pi^2$:
\begin{eqnarray}
\frac{m_{c\bar{c}}}{m_{\Omega}}=\frac{m^0_{c\bar{c}}}{m_{\Omega}^0}+\frac{m_\pi^2}{4\pi f_\pi m_{\Omega}}\left(\sigma_{c\bar{c}}-\frac{m_{c\bar{c}}{\sigma_{\Omega}}}{m_\Omega}\right)+\mathcal{O}(m_\pi^4),
\end{eqnarray}
where $m^0_{c\bar{c}}$ is the bare charmonium mass, and the $a$-dependence is parametrized by $c_a (m^{\rm{phys}}_\Omega a)^2$.
\begin{center}
\begin{table}
\begin{tabular}{|c|ccc|}
\hline
Hadron&$^{2S+1}L_J$&$J^{PC}$&Interpolator
\\\hline \hline
    $ \eta_{{c}}$&$^1S_0$&$0^{-+}$&${\bar Q}^k_{c}\gamma_5Q^k_{c}$  \\
     $(J/\psi)^i$&$^3S_1$&$1^{--}$&${\bar Q}^k_{c}\gamma^iQ^k_{c}$ \\
 $\chi_{c0}$&$^1P_0$&$0^{++}$&${\bar Q}^k_{c}Q^k_{c}$ \\
  $(\chi_{c1})^i$&$^3P_1$&$1^{++}$&${\bar Q}^k_{c}\gamma_5\gamma^iQ^k_{c}$ \\
  $(h_{c})^{ji}$&$^3P_1$&$1^{+-}$&${\bar Q}^k_{c}\gamma^j\gamma^iQ^k_{c}$
\\\hline
\hline
\end{tabular}

\caption{Interpolating operators for the charmonium spectrum. ${{Q}}^k_{c}$ labels the the charm quark with color index $k$. }
\label{cc_oper}

\end{table}
\end{center}
\begin{center}
\begin{table}
\begin{tabular}{c|cccccc}
\hline
Hadron& $m_c$& $(a{m_{{H}}})_{\rm A1}$& $(a{m_{{H}}})_{\rm A2}$& $(a{m_{{H}}})_{\rm B1}$& $(a{m_{{H}}})_{\rm B2}$& $(a{m_{{H}}})_{\rm C1}$  \\\hline
\hline
\multirow{2}{*}{$\overline{1S}$}&$m_{c1}$&1.86213(61)(21)
[8--16]&1.85571(32)(5)
[9--15]&1.37703(27)(18)
[11--18]&1.36696(41)(26)
[13--23]&0.93723(18)(11)
[12--32]\\
&$m_{c2}$&1.83438(47)(24)
[8--16]&1.80666(32)(5)
[9--15]&1.34397(28)(20)
[11--18]&1.32897(41)(15)
[13--23]&0.92157(18)(9)
[12--32]\\
\hline
\multirow{2}{*}{$\eta_c$}&$m_{c1}$&1.86213(61)(19)
[7--13]&1.85571(32)(4)
[16--24]&1.37703(27)(15)
[24--36]&1.36696(41)(39)
[20--25]&0.93723(18)(14)
[18--36]\\
&$m_{c2}$&1.83438(47)(27)
[7--13]&1.80666(32)(4)
[16--24]&1.34397(28)(20)
[24--36]&1.32897(41)(13)
[20--25]&0.92157(18)(16)
[18--36]\\
\hline
\multirow{2}{*}{$J/\psi$}&$m_{c1}$&1.91025(50)(29)
[5--17]&1.90212(54)(21)
[13--17]&1.41634(78)(19)
[26--30]&1.40612(53)(18)
[21--26]&0.96470(29)(38)
[18--36]\\
&$m_{c2}$&1.88354(44)(30)
[5--17]&1.85446(55)(15)
[13--17]&1.38428(81)(29)
[26--30]&1.36975(54)(9)
[21--26]&0.94955(30)(22)
[18--36]\\
\hline
\multirow{2}{*}{$\chi_{c0}$}&$m_{c1}$&2.1382(22)(19)
[4--9]&2.1264(23)(34)
[5--12]&1.5873(29)(27)
[7--10]&1.5599(44)(22)
[12--23]&1.0619(34)(20)
[17--22]\\
&$m_{c2}$&2.1126(19)(17)
[4--9]&2.0787(23)(28)
[5--12]&1.5537(28)(25)
[7--10]&1.5209(46)(21)
[12--23]&1.0557(19)(20)
[17--22]\\
\hline
\multirow{2}{*}{$\chi_{c1}$}&$m_{c1}$&2.164(11)(5)
[11--16]&2.1574(56)(36)
[9--12]&1.6121(26)(12)
[3--9]&1.6001(53)(31)
[12--23]&1.0966(37)(16)
[12--18]\\
&$m_{c2}$&2.133(10)(4)
[11--16]&2.1104(57)(44)
[9--12]&1.5807(26)(16)
[3--9]&1.5631(55)(33)
[12--23]&1.0814(39)(10)
[12--18]\\
\hline
\multirow{2}{*}{$h_c$}&$m_{c1}$&2.1612(93)(60)
[7--10]&2.1573(54)(62)
[9--14]&1.6296(59)(45)
[10--17]&1.6078(59)(45)
[12--23]&1.0904(89)(36)
[18--24]\\
&$m_{c2}$&2.1373(90)(65)
[7--10]&2.1105(55)(35)
[9--14]&1.5952(83)(74)
[10--17]&1.5709(59)(47)
[12--23]&1.0869(53)(20)
[18--24]\\
\hline
\multirow{2}{*}{$D_s$}&$m_{c1}$&
1.20785(70)(38)[13--23]
&
1.20348(65)(29)[8--15]
&
0.89883(46)(40)[13--29]
&
0.88914(62)(46)[13--23]
&
0.61196(53)(37)[18--26]
\\
&$m_{c2}$&
1.19203(69)(40)[13--23]
&
1.17734(64)(26)[8--15]
&
0.88112(45)(36)[13--29]
&
0.86803(59)(45)[13--23]
&
0.60333(52)(28)[18--26]\\
\hline

\hline
\hline
 \end{tabular}
\caption{Charmonium and $D_s$ masses in lattice units for the five ensembles and two charm masses. Errors listed are statistical and fitting window systematic. The fitting window is given in square brackets.  }
\label{cc_ens}
\end{table}
\end{center}

\begin{figure}
\begin{center}
\includegraphics[totalheight=12cm]{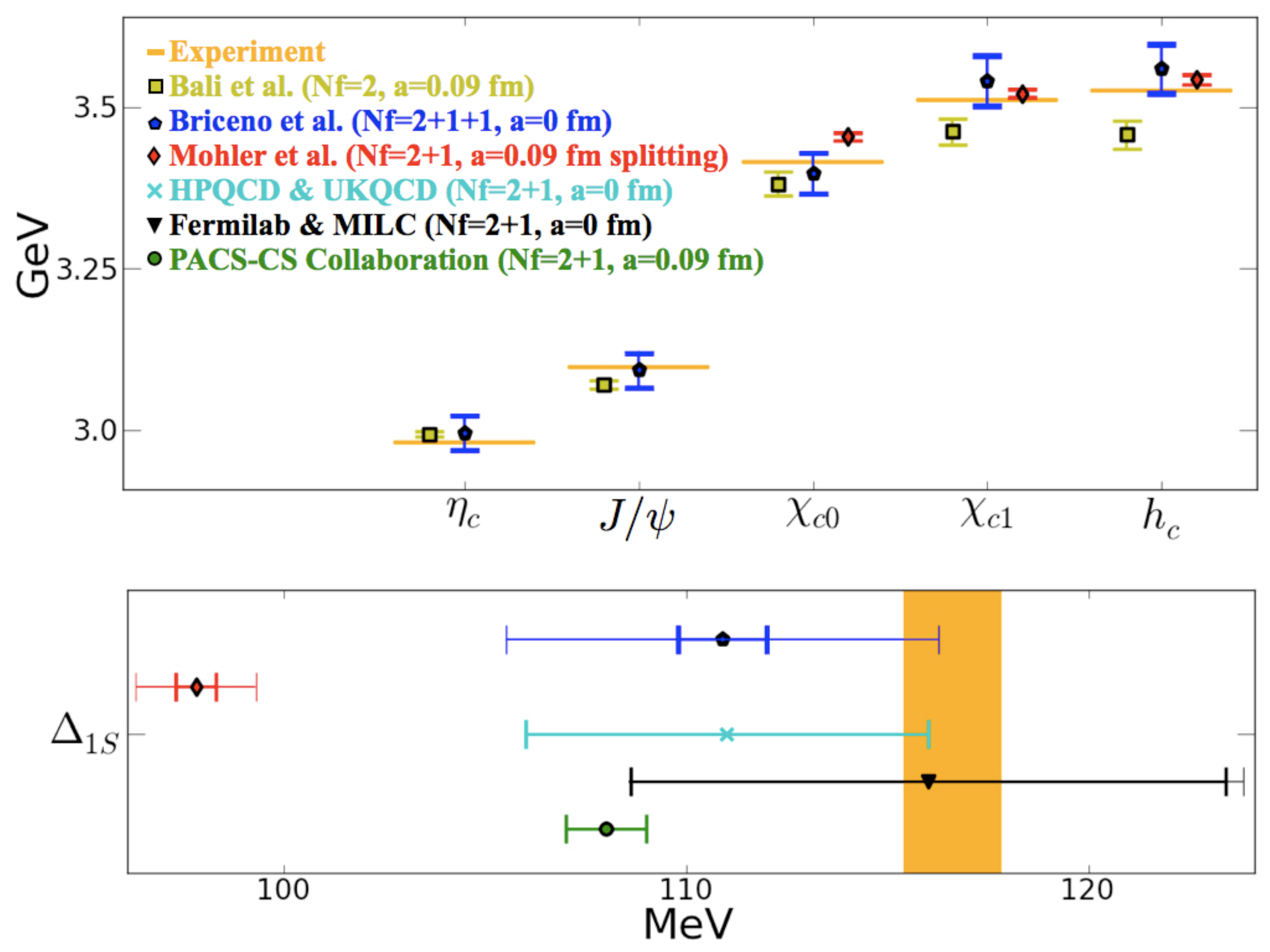}
\end{center}
\caption{Our determination of the low-lying charmonium spectrum after extrapolating to the physical point, labeled as ``Briceno~et~al.'', as well as a survey of previous unquenched lattice calculations~\cite{meson1, disc3, meson3, meson4, meson5}. Mohler~et~al. determined the spitting between $\{\chi_{c0},\chi_{c1},h_c\}$ and $\overline{1S}$~\cite{meson1}; in order to compare their results with ours, we have set $\overline{1S}$ to its physical value, while leaving their hyperfine splitting unchanged. The statistical uncertainty is shown as a thick inner error bar, while the statistical and systematic uncertainties (if estimated in the paper) added in quadrature are shown as a larger thin error bar. Our systematic uncertainties include errors originating from the fitting window, scale setting, pion mass determination, finite-volume effects, $\mathcal{O}(m^4_\pi,a^2m_\pi)$-corrections to the expressions used to extrapolate to the physical point, and the strange mass tuning (as discussed in Sec.~\ref{system}). The orange line/band indicates the experimentally measured masses or splittings with their corresponding uncertainties~\cite{pdg}. }\label{charmonium}
\end{figure}

\begin{figure}
\begin{center}
\includegraphics[totalheight=5cm]{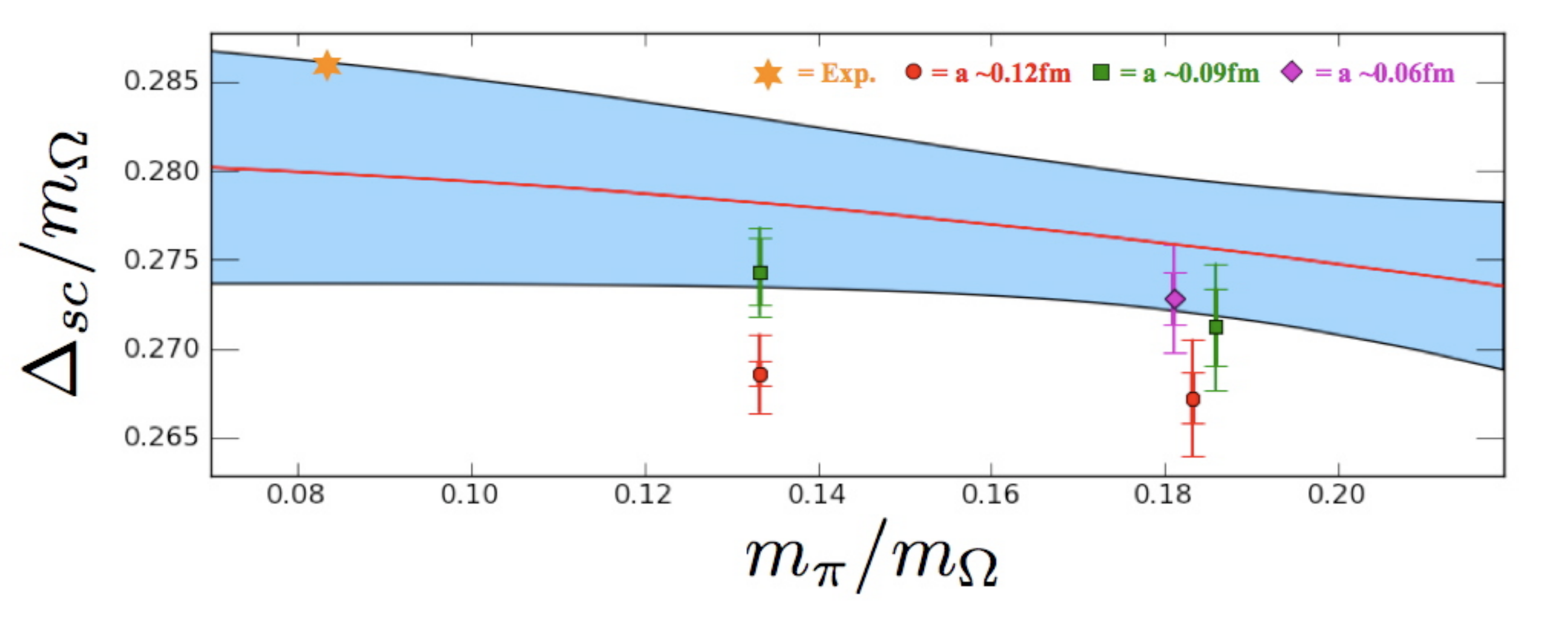}
\end{center}
\caption{$\chi$PT and continuum extrapolations of the $\Delta_{sc}=m_{D_s}-m_{\eta_c/2}$ splitting. The red line indicates the fit of the data that has been extrapolated to $a=0$. The blue band include the statistical and systematic errors added in quadrature.}\label{deltasc}
\end{figure}

Using this procedure, we have verified that our calculations reproduce the experimental low-lying charmed-meson spectrum. In Fig.~\ref{charmonium} we show our results for the charmonium spectrum (as well as the hyperfine splitting $\Delta_{1S}\equiv M_{J/\psi}-M_{\eta_c}$) after extrapolating to the physical point. As a result, our errorbars are larger than those of other calculations. For comparison, we show in Fig.~\ref{charmonium} a sample of previous dynamical lattice calculations that have studied the charmonium spectrum. By comparing the level of precision of $am_\Omega$ (see Table~\ref{ensembles2}) and $am_{c\bar{c}}$ (see Table~\ref{cc_ens}), one can see that it is the uncertainty of $am_\Omega$ that dominates the overall uncertainty of the $m_{c\bar{c}}/m_\Omega$ ratio.

The works by Bali~et~al. and Mohler~et~al. are far more extensive than the small sample that is being represented here. Both groups used the variational method over different sources and sinks to not only extract ground-state energies but also those of the excited states. Mohler~et~al. evaluated the spectrum for the $\{c\bar{c},c\bar{s},c\bar{l}\}$ systems for a range of six pion masses ranging from 702~MeV to 156~MeV at a single lattice spacing, $a\approx0.09$~fm. On the other hand, Bali~et~al. evaluated the $\{c\bar{c}\}$ spectrum, including disconnected diagrams, at three lattice spacings but did not provide a continuum-extrapolated result for the spectrum or an estimate of the discretization error.

The conclusion of Fig.~\ref{charmonium} is evident: these non-continuum results come with a large systematic error due to nonzero lattice spacing. This error decreases with lattice spacing, but from Fig.~\ref{charmonium} it is clear that in order to reproduce the physical spectrum, it is necessary to extrapolate masses to the continuum. For example, in the upper figure in Fig.~\ref{charmonium} we see that despite our masses having the largest uncertainties, ours are the only results that are consistently in agreement with experiment. We conclude that previous calculations that do not extrapolate their results to the continuum have underestimated their systematic errors.

When tuning the charm mass to the spin-averaged mass, $\overline{1S}$, the most natural quantity to study is the hyperfine splitting $\Delta_{1S}$. As a result, this splitting has received a great deal of attention in the community. One surprising feature is that for a finite lattice spacing, $\Delta_{1S}$ is underestimated~\cite{meson1, meson4}. In our calculations we find the value of $\Delta_{1S}$ agrees with experiment only after extrapolating to the continuum. This is consistent with the findings of the HPQCD/UKQCD Collaboration~\cite{meson1} and Fermilab Lattice and MILC Collaborations~\cite{meson5}, as shown in the lower part of Fig.~\ref{charmonium}. Therefore, it cannot be overstated that charmed quantities need to be evaluated at multiple lattice spacings to properly quantify the systematics.

In order to further test the strange- and charm-mass tuning, we evaluated the $\Delta_{sc}\equiv m_{D_s}-m_{\eta_c/2}$ splitting. This is the binding-energy difference between the heavy-light and heavy-heavy systems; there is no reliable analytical procedure for calculating this quantity.  Since the strange-charm meson, $D_s$, has no light degrees of freedom, up to $\mathcal{O}(m_\pi^3)$ its mass is linear in $m_\pi^2$, therefore the $\Delta_{sc}$ splitting can be extrapolated using
\begin{eqnarray}
\frac{{\Delta_{sc}}}{m_{\Omega}}=\frac{\Delta^0_{sc}}{m_{\Omega}^0}+\frac{m_\pi^2}{4\pi f_\pi m_{\Omega}}\left(\sigma_{\Delta_{sc}}-\frac{{\sigma_{\Omega}\Delta_{sc}}}{m_\Omega}\right)+\mathcal{O}(m_\pi^4).
\end{eqnarray}
where $\Delta^0_{sc}$ denotes the bare splitting, and we extrapolate to the continuum using $c_a (m^{\rm{phys}}_\Omega a)^2$.

In Table~\ref{cc_ens}, the $D_s$ and $\eta_c$ meson masses are shown for each ensemble. Figure ~\ref{deltasc} shows the values for the $\Delta_{sc}$ splitting after continuum extrapolation, along with their corresponding statistical and systematic uncertainties (see Sec.~\ref{system}). Figure~\ref{deltasc} shows that the $a$-dependence of $\Delta_{sc}$ is sizable; in fact, continuum extrapolation is necessary in order to reproduce the physical value. In performing the continuum extrapolation of $\Delta_{sc}$, we find the $a$-dependent LEC to be $c_a=-0.0088(46)$. Since our determination of the $c\bar{c}$ and $c\bar{s}$ spectrum is in agreement with experiment, we believe that the estimates of the systematics in Sec.~\ref{system} accurately reflect the sources of systematic error of the calculation presented in this paper. 


\begin{figure}
\begin{center}
\includegraphics[totalheight=9cm]{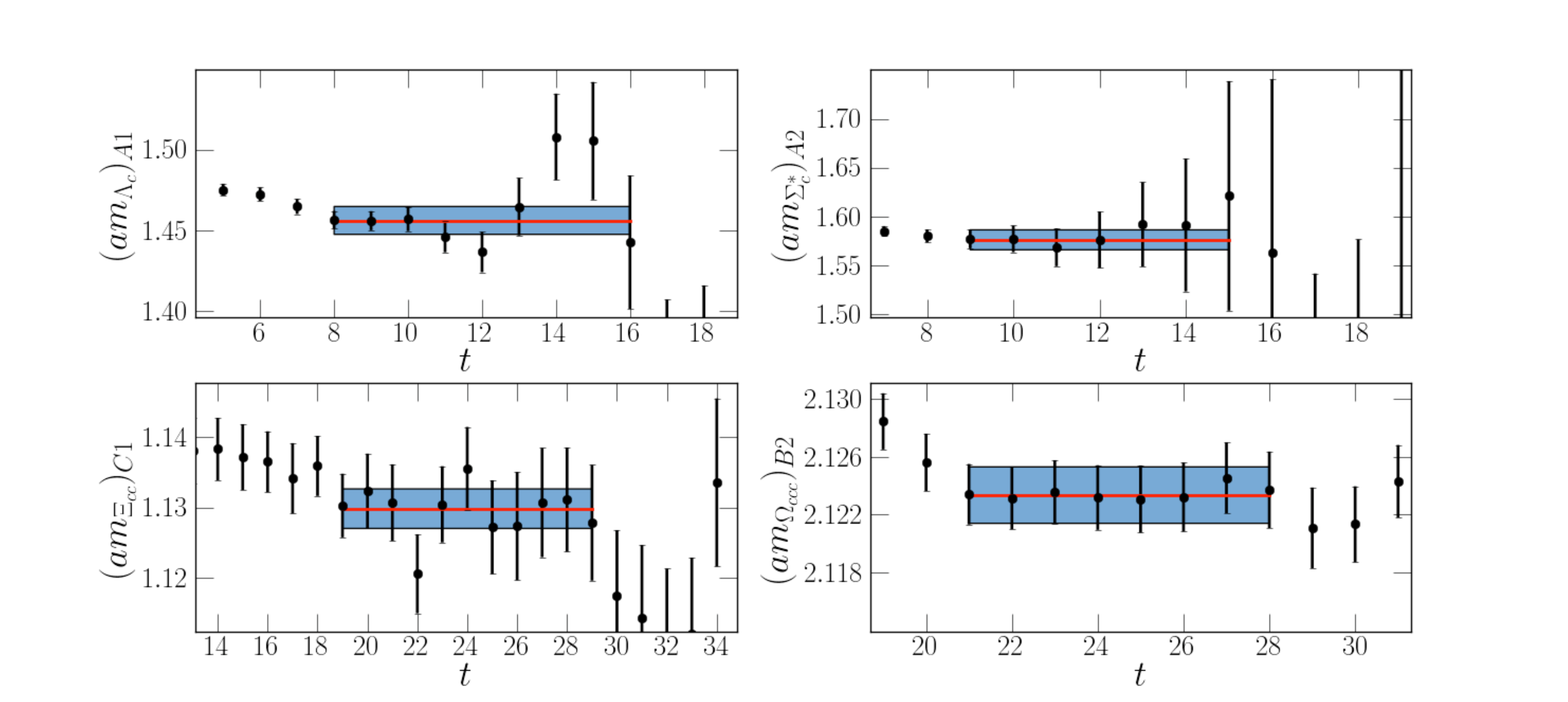}
\end{center}
\caption{Sample effective-mass plots from the various ensembles of the charmed-baryon sector. The errorbar shown includes the statistical and systematic uncertainty (from varying the fitted range) added in quadrature. }\label{EMPs}
\end{figure}
\begin{center}
\begin{table}
\begin{tabular}{|c|c|}
\hline
$J^P=\frac{1}{2}^+$ & $J^P=\frac{3}{2}^+$  \\\hline \hline
    $ \Lambda_{{c}}=\epsilon^{{klm}}{P}^+{Q}^{{k}}_{{c}}\left({q}^{{lT}}_{{u}}\Gamma^{{A}}{q}^{m}_{{d}}\right)$  &   \\
$\Xi_{{c}}=\epsilon^{{klm}}{P}^+{Q}^{{k}}_{{c}}\left({q}^{{lT}}_{{u}}\Gamma^{{A}}{q}^{m}_{{s}}\right)$
&
$(\Sigma_{{c}}^{*})^{{i}}=
\epsilon^{{klm}}{P}^+({P}^{3/2}_{{E}})^{{ij}}{Q}^{{k}}_{{c}}\left({q}^{{lT}}_{{u}}\Gamma^{{j}}{q}^{m}_{{u}}\right),$\\
$ (\Sigma_{{c}})^{{i}}=
\epsilon^{{klm}}{P}^+({P}^{1/2}_{{E}})^{{ij}}{Q}^{{k}}f_{{c}}\left({q}^{{lT}}_{{u}}\Gamma^{{j}}{q}^{m}_{{u}}\right)$
&$(\Xi_{{c}}^{*})^{{i}}=\frac{\epsilon^{{klm}}}{\sqrt{2}}
{P}^+({P}^{3/2}_{{E}})^{{ij}}{Q}^{{k}}_{{c}}\left({q}^{{lT}}_{{u}}\Gamma^{{j}}{q}^{m}_{{s}}
+{q}^{{lT}}_{{s}}\Gamma^{{j}}{q}^{m}_{{u}}\right)$\\
$(\Xi'_{{c}})^{{i}}=\frac{\epsilon^{{klm}}}{\sqrt{2}}
{P}^+({P}^{1/2}_{{E}})^{{ij}}{Q}^{{k}}_{{c}}\left({q}^{{lT}}_{{u}}\Gamma^{{j}}{q}^{m}_{{s}}
+{q}^{{lT}}_{{s}}\Gamma^{{j}}{q}^{m}_{{u}}\right)$
&$(\Omega_{{c}}^{*})^{{i}}=
\epsilon^{{klm}}{P}^+({P}^{3/2}_{{E}})^{{ij}}{Q}^{{k}}_{{c}}\left({q}^{{lT}}_{{s}}\Gamma^{{j}}{q}^{m}_{{s}}\right)$
\\
$(\Omega_{{c}})^{{i}}=
\epsilon^{{klm}}{P}^+({P}^{1/2}_{{E}})^{{ij}}{Q}^{{k}}_{{c}}\left({q}^{{lT}}_{{s}}\Gamma^{{j}}{q}^{m}_{{s}}\right)$
&$(\Xi_{{cc}}^{*})^{{i}}=
\epsilon^{{klm}}{P}^+({P}^{3/2}_{{E}})^{{ij}}{q}^{{k}}_{{u}}\left({Q}^{{lT}}_{{c}}\Gamma^{{j}}{Q}^{m}_{{c}}\right)$\\
$(\Xi_{{cc}})^{{i}}=
\epsilon^{{klm}}{P}^+({P}^{1/2}_{{E}})^{{ij}}{q}^{{k}}_{{u}}\left({Q}^{{lT}}_{{c}}\Gamma^{{j}}{Q}^{m}_{{c}}\right)$
&$(\Omega_{{cc}}^{*})^{{i}}=
\epsilon^{{klm}}{P}^+({P}^{3/2}_{{E}})^{{ij}}{q}^{{k}}_{{s}}\left({Q}^{{lT}}_{{c}}\Gamma^{{j}}{Q}^{m}_{{c}}\right)$\\

$(\Omega_{{cc}})^{{i}}=
\epsilon^{{klm}}{P}^+({P}^{1/2}_{{E}})^{{ij}}{q}^{{k}}_{{s}}\left({Q}^{{lT}}_{{c}}\Gamma^{{j}}{Q}^{m}_{{c}}\right)$
&$(\Omega_{{ccc}})^{{i}}=
\epsilon^{{klm}}{P}^+({P}^{3/2}_{{E}})^{{ij}}{Q}^{{k}}_{{c}}\left({Q}^{{lT}}_{{c}}\Gamma^{{j}}{Q}^{m}_{{c}}\right)$\\\hline
\hline
\end{tabular}
\caption{The interpolating operators for the positive-parity baryons~\cite{UKQCD}. ${q}_{u,d,s}$ respectively denote the up-, down- and strange-quark annihilation operators, ${Q}_{c}$ denotes the charm-quark operator, $\{{k,l,m}\}$ are color indices, while $\{{i,j}\}$ denote polarization indices. $\left(\Gamma^{{A}},\Gamma^{{i}}\right)$ are the antisymmetric and symmetric spin matrices $({C}\gamma_5, {C}\gamma^i)$, where $C$ is the charge-conjugation matrix. In order to have the best possible overlap with the state of interest, we have used the spin projection operators $({P}^{3/2}_{E})^{ij}= \delta^{ij}-\frac{1}{3}\gamma^{i}\gamma^{j}$ and $( {P}^{1/2}_{E})^{ij}= \delta^{ij}-({P}^{3/2}_{E})^{ij}=\frac{1}{3}\gamma^{i}\gamma^{j}$, and the positive-parity projection operator ${P}^+={(1+\gamma^4)}/{2}$.}
\end{table}
\label{baroper}
\end{center}

 \section{Charmed-Baryon Spectrum \label{cbs}}

With confidence that our tuning reproduces the low-lying $c\bar{c}, c\bar{s}, l\bar{s}$ spectrum within our systematics, we proceed to evaluate the positive-parity charmed-baryon spectrum. Heavy-quark symmetry dictates that the quantum numbers of the light degrees of freedom of any heavy-light system are conserved. One can identify approximately degenerate multiplets by these quantum numbers. For singly charmed baryons, the light degrees of freedom can have total spin equal to zero or one. Under $SU(3)_V$ chiral symmetry, the spin-singlet multiplet transforms as a $\bar{\mathbf{3}}$ irrep. The spin triplet is a $\mathbf{6}$ irrep when the total angular momentum is $J=1/2$ and a $\mathbf{6}^*$ irrep when the total angular momentum is $J=3/2$. In the heavy-quark limit, these are degenerate. The doubly charmed baryons form a $\mathbf{3}$ irrep when the total angular momentum is $J=1/2$ and a $\mathbf{3}^*$ irrep when the total angular momentum is $J=3/2$. The triply charmed baryons are singlets under $SU(3)_V$. This algebra was manifested by the interpolating operators used in this calculation, as shown in Table~\ref{baroper}~\cite{UKQCD}.
Figure~\ref{EMPs} displays examples of the effective-mass plots for various correlation functions. Table~\ref{baryon_ens} lists the baryon masses in lattice units for each charm-quark mass and ensemble along with the statistical and fitting-window systematic uncertainties and the chosen fitting window.

\begin{center}
\begin{table}
\begin{tabular}{c|cccccc}
\hline
\hline
Hadron& $m_c$& $(am_{H})_{A1}$& $(am_{H})_{A2}$& $(am_{H})_{B1}$& $(am_{H})_{B2}$& $(am_{H})_{C1}$  \\
\hline
\multirow{2}{*}{$\Lambda_c$}&$m_{c1}$&1.4561(42)(70)[8--16]&1.4228(77)(73)[9--15]&1.0808(42)(33)[11--15]&1.0328(102)(79)[16--25]&0.7339(56)(15)[19--22]\\
&$m_{c2}$&1.4401(42)(70)[8--16]&1.3976(76)(69)[9--15]&1.0643(41)(35)[11--15]&1.0136(98)(62)[16--25]&0.7258(56)(15)[19--22]\\
\hline
\multirow{2}{*}{$\Xi_c$}&$m_{c1}$&1.5333(24)(28)[8--15]&1.5120(31)(20)[8--15]&1.1438(37)(21)[14--18]&1.1115(37)(47)[14--27]&0.7747(48)(10)[26--29]\\
&$m_{c2}$&1.5174(24)(27)[8--15]&1.4871(31)(21)[8--15]&1.1274(37)(19)[14--18]&1.0922(33)(20)[14--27]&0.7665(48)(10)[26--29]\\
\hline
\multirow{2}{*}{$\Sigma_c$}&$m_{c1}$&1.5521(40)(30)[8--11]&1.5286(50)(54)[8--16]&1.1703(43)(25)[11--16]&1.1351(80)(78)[13--23]&0.7968(32)(54)[13--22]\\
&$m_{c2}$&1.5359(40)(30)[8--11]&1.5028(50)(51)[8--16]&1.1530(43)(29)[11--16]&1.1134(74)(52)[13--23]&0.7883(32)(54)[13--22]\\
\hline
\multirow{2}{*}{$\Sigma^*_c$}&$m_{c1}$&1.6178(43)(48)[7--11]&1.5760(91)(44)[9--15]&1.1979(83)(53)[13--19]&1.1731(105)(167)[13--20]&0.8055(83)(29)[19--24]\\
&$m_{c2}$&1.6020(43)(50)[7--11]&1.5516(91)(42)[9--15]&1.1812(82)(55)[13--19]&1.1569(97)(76)[13--20]&0.7975(83)(29)[19--24]\\
\hline
\multirow{2}{*}{$\Xi'_c$}&$m_{c1}$&1.5878(60)(78)[12--23]&1.5820(55)(54)[11--18]&1.1925(51)(14)[16--22]&1.1682(49)(34)[15--21]&0.8089(23)(22)[12--23]\\
&$m_{c2}$&1.5717(60)(86)[12--23]&1.5564(55)(51)[11--18]&1.1753(50)(15)[16--22]&1.1471(44)(16)[15--21]&0.8005(22)(23)[12--23]\\
\hline
\multirow{2}{*}{$\Xi^*_c$}&$m_{c1}$&1.662(3)(14)[8--18]&1.6388(58)(41)[10--14]&1.2314(65)(41)[15--21]&1.2060(54)(48)[14--21]&0.8328(54)(17)[20--24]\\
&$m_{c2}$&1.646(3)(14)[8--18]&1.6142(57)(39)[10--14]&1.2157(64)(39)[15--21]&1.1896(51)(9)[14--21]&0.8248(53)(17)[20--24]\\
\hline
\multirow{2}{*}{$\Omega_c$}&$m_{c1}$&1.6487(69)(16)[16--24]&1.6393(22)(24)[8--14]&1.2280(45)(17)[19--23]&1.2129(28)(3)[15--19]&0.8341(25)(25)[18--24]\\
&$m_{c2}$&1.6322(69)(16)[16--24]&1.6138(22)(24)[8--14]&1.2112(45)(16)[19--23]&1.1919(25)(3)[15--19]&0.8262(24)(24)[18--24]\\
\hline
\multirow{2}{*}{$\Omega^*_c$}&$m_{c1}$&1.6960(38)(52)[11--20]&1.6882(27)(29)[8--14]&1.2567(64)(34)[19--26]&1.2493(32)(17)[14--19]&0.8567(24)(24)[15--30]\\
&$m_{c2}$&1.6805(38)(52)[11--20]&1.6638(27)(28)[8--14]&1.2408(64)(29)[19--26]&1.2313(29)(7)[14--19]&0.8489(23)(25)[15--30]\\
\hline
\multirow{2}{*}{$\Xi_{cc}$}&$m_{c1}$&2.2349(33)(42)[11--25]&2.2194(67)(61)[15--22]&1.6628(21)(13)[6--16]&1.6413(46)(17)[17--25]&1.1298(25)(12)[19--29]\\
&$m_{c2}$&2.2037(33)(39)[11--25]&2.1701(66)(56)[15--22]&1.6394(48)(50)[6--16]&1.6070(39)(21)[17--25]&1.1139(25)(12)[19--29]\\
\hline
\multirow{2}{*}{$\Xi^*_{cc}$}&$m_{c1}$&2.3053(26)(27)[8--16]&2.2455(115)(72)[15--19]&1.6381(55)(47)[18--26]&1.6801(66)(37)[17--22]&1.1570(91)(32)[32--41]\\
&$m_{c2}$&2.2744(25)(27)[8--16]&2.1970(114)(73)[15--19]&1.6808(29)(44)[18--26]&1.6459(56)(27)[17--22]&1.1416(91)(34)[32--41]\\
\hline
\multirow{2}{*}{$\Omega_{cc}$}&$m_{c1}$&2.2893(28)(9)[17--26]&2.2739(22)(12)[15--27]&1.7008(18)(2)[18--26]&1.6786(33)(14)[24--28]&1.1562(14)(4)[19--29]\\
&$m_{c2}$&2.2580(28)(10)[17--26]&2.2247(21)(12)[15--27]&1.6677(18)(3)[18--26]&1.6417(28)(6)[24--28]&1.1403(14)(4)[19--29]\\
\hline
\multirow{2}{*}{$\Omega^*_{cc}$}&$m_{c1}$&2.3385(66)(29)[11--19]&2.3178(31)(19)[15--23]&1.7331(43)(10)[23--29]&1.7180(38)(23)[20--26]&1.1796(21)(6)[26--31]\\
&$m_{c2}$&2.3078(66)(29)[11--19]&2.2694(31)(19)[15--23]&1.7001(43)(9)[23--29]&1.6799(35)(16)[20--26]&1.1641(21)(6)[26--31]\\
\hline
\multirow{2}{*}{$\Omega_{ccc}$}&$m_{c1}$&2.9621(16)(9)[16--24]&2.9466(15)(17)[16--24]&2.1953(15)(7)[32--39]&2.1788(18)(2)[21--28]&1.4921(22)(8)[38--43]\\
&$m_{c2}$&2.9161(16)(8)[16--24]&2.8753(15)(17)[16--24]&2.1472(16)(8)[32--39]&2.1239(17)(2)[21--28]&1.4690(23)(4)[38--43]\\
\hline \end{tabular}
\caption{Charmed-baryon masses for the five ensembles in lattice units, statistical and fitting window systematic uncertainties, and fitting windows.}
\label{baryon_ens}
\end{table}
\end{center}



\subsection{Chiral and Continuum Extrapolation}
\begin{figure}
\begin{center}
\includegraphics[totalheight=3cm]{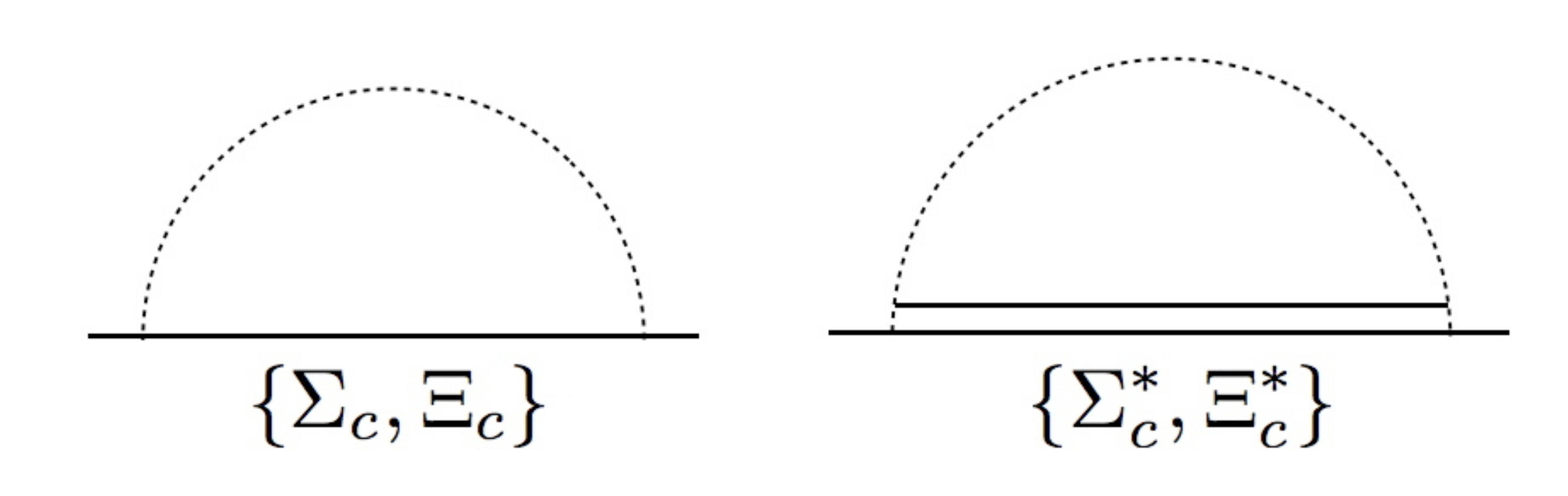}
\end{center}
\caption{Two of the self-energy diagrams contributing to the masses of a singly charmed baryon in the $\textbf{6}$ irrep. The first depicts contributions arising from loops containing a pion and a member of the $\textbf{6}$ irrep, while the second correspond to loops containing a pion and a member of the $\textbf{6}^*$ irrep. There are similar self-energy diagrams for baryons in the $\textbf{6}^*$ irrep.}\label{se}
\end{figure}

As discussed in Sec.~\ref{hqaction}, the ratios of each charmed-hadron mass to the $\Omega$ mass are interpolated to the physical charm mass, defined by ${m_{\overline{1S}}}/{m_{\Omega}}= 1.83429(56)$. After this is done for each ensemble, it is necessary to extrapolate the ratios to the physical light-quark mass and continuum. Due to the rather large expansion parameter of $SU(3)$ $\chi$PT and poorer convergence rate, we use $SU(2)$ HH$\chi$PT to extrapolate the baryon masses to the physical pion mass. Previous HH$\chi$PT calculations of the singly charmed-baryon masses used the static limit, $m_Q\rightarrow \infty$~\cite{hhchipt6, hhchipt7}. At $\mathcal{O}(1/m_Q)$ new operators are introduced that explicitly break the $\bold 6$-$\bold 6^*$ degeneracy~\cite{hhchipt5}, resulting in three independent bare splittings $\{\Delta_{\bar{\bold 3},\bold 6},\Delta_{\bar{\bold 3},\bold 6^*},\Delta_{{\bold 6},\bold 6^*}\}$. We extend previous work to include the $\mathcal{O}(1/m_Q)$ corrections for the $\{\Lambda_c,\Sigma_c,\Sigma^*_c\}$ and $\{\Xi_c,\Xi'_c,\Xi^*_c\}$ multiplets by evaluating the contribution arising from the two self-energy diagrams depicted in Fig.~\ref{se}.

First, consider the $\{\Lambda_c,\Sigma_c,\Sigma^*_c\}$ multiplet. Up to $\mathcal{O}(m_\pi^3)$, the $m_\pi$ dependence of the ratio of the particle masses to $m_\Omega$ can be written as
\begin{eqnarray}
\label{lambda}
\frac{m_{\Lambda_c}}{m_\Omega}&=&\frac{m^0_{\Lambda_c}}{m_\Omega^0}+\frac{\bar\sigma_{\Lambda_c}m^2_{\pi}}{(4\pi f_\pi)m_\Omega}
-\frac{6g_3^2}{{(4\pi f_\pi)^2m_\Omega}}\left(\frac{1}{3}\mathcal{F}(m_\pi,\Delta_{\Lambda_c\Sigma_c},\mu)+
\frac{2}{3}\mathcal{F}(m_\pi,\Delta_{\Lambda_c\Sigma^*_c},\mu)\right)
\\
\frac{m_{\Sigma_c}}{m_\Omega}&=&\frac{m^0_{\Lambda_c}+\Delta^0_{\Lambda_c\Sigma_c}}{m_\Omega^0}
+\frac{\bar\sigma_{\Sigma_c}m^2_{\pi}}{(4\pi f_\pi)m_\Omega}
-\frac{2g_3^2}{{3(4\pi f_\pi)^2m_\Omega}} \mathcal{F}(m_\pi,-\Delta_{\Lambda_c\Sigma_c},\mu)
+\frac{g_2^2}{{(4\pi f_\pi)^2m_\Omega}}\left(\frac{4}{9}\mathcal{F}(m_\pi,0,\mu)+
\frac{8}{9}\mathcal{F}(m_\pi,\Delta_{\Sigma_c\Sigma^*_c},\mu)\right)
\nn\\
\frac{m_{\Sigma^*_c}}{m_\Omega}&=&\frac{m^0_{\Lambda_c}+\Delta^0_{\Lambda_c\Sigma^*_c}}{m_\Omega^0}
+\frac{\bar\sigma_{\Sigma^*_c}m^2_{\pi}}{(4\pi f_\pi)m_\Omega}
-\frac{2g_3^2}{{3(4\pi f_\pi)^2m_\Omega}}\mathcal{F}(m_\pi,-\Delta_{\Lambda_c\Sigma^*_c},\mu)
+\frac{g_2^2}{{(4\pi f_\pi)^2m_\Omega}}\left(\frac{10}{9}\mathcal{F}(m_\pi,0,\mu)+
\frac{2}{9}\mathcal{F}(m_\pi,-\Delta_{\Sigma_c\Sigma^*_c},\mu)\right),\nn
\end{eqnarray}
where $\bar{\sigma}_{H}=\left(\sigma_{H}-{m^0_{H}{\sigma_{\Omega}}}/{m_\Omega}\right)$, $m^0$ and $\Delta^0$ label the bare masses and splittings, and $g$'s and $\sigma$'s are the LECs of the theory. The chiral function $\mathcal{F}$ is defined as
\begin{eqnarray}
\mathcal{F}(m,\Delta,\mu)=(\Delta^2-m^2+i\epsilon)^{3/2}\ln\left(\frac{\Delta+ \sqrt{\Delta^2-m^2+i\epsilon}}{\Delta- \sqrt{\Delta^2-m^2+i\epsilon}}\right)-\frac{3}{2}\Delta m^2\ln\left(\frac{m^2}{\mu^2}\right)-\Delta^3\ln\left(\frac{4\Delta^2}{m^2}\right),
\end{eqnarray}
with $\mathcal{F}(m,0,\mu)=\pi m_\pi^3$. From Eq.~\ref{lambda}, in the static limit we reproduce the previous results~\cite{hhchipt6, hhchipt7}. For the extrapolation to the continuum limit, we consider the lattice-spacing dependence of $\delta_r(a)={c_a} (m^{\rm{phys}}_\Omega a)^2$ for each baryon within the same multiplet to have the same behavior.

\begin{figure}
\begin{center}
\includegraphics[totalheight=10cm]{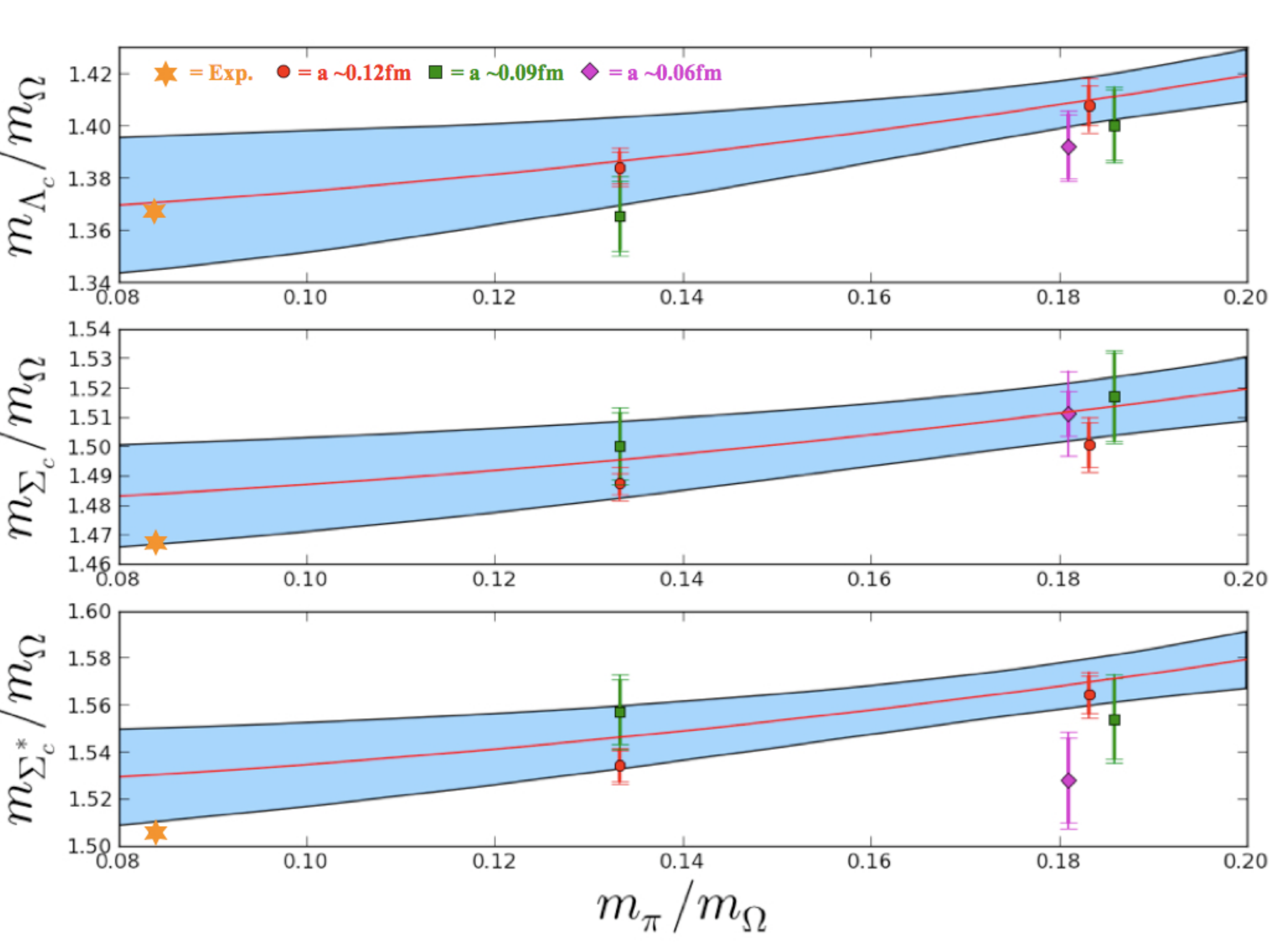}
\end{center}
\caption{NLO HH$\chi$PT and continuum simultaneous extrapolations of $\{\Lambda_c,\Sigma_c,\Sigma^*_c\}$ masses. The red line depicts the fit of the data that has been extrapolated to $a=0$. The blue band includes the statistical and systematic errors added in quadrature. }
\label{lambda_plots}
\end{figure}

In order to stabilize our fits, we evaluate the splittings $\{\Delta_{\Lambda_c\Sigma_c},\Delta_{\Sigma_c\Sigma_c^*},\Delta_{\Lambda_c \Sigma_c^*}\}$ for each ensemble and extrapolate them to the physical pion mass with the assumption that their lattice-spacing dependence is suppressed. The resulting splittings serve as input to the chiral function in Eq.~\ref{xicc}. In addition, when minimizing $\chi^2$ we require the axial couplings to be real, $g^2>0$. This requirement assures that the HH$\chi$PT Lagrangian is Hermitian, and it reduces the parameter space of the minimization routine, thereby resulting in smaller uncertainties while leaving the mean values of the extrapolated masses unchanged.
The scale $\mu$ is set to $700$~MeV; we do not observe a difference in the results when $\mu$ is varied among $\{600\mbox{ MeV}, 700\mbox{ MeV}, 800\mbox{ MeV}\}$.
Using the physical value of $m^{\rm{phys}}_\Omega/{f^{\rm{phys}}_\pi}=12.796(37)$, we find the LECs shown in Table~\ref{lambda_lec}.
In Fig.~\ref{lambda_plots} we display our fits at the continuum ($a=0$) along with the value of $m_H/m_\Omega$ for each ensemble as a function of $m_\pi/m_\Omega$. From Fig.~\ref{lambda_plots}, one sees all masses are within $1.1~\sigma$ of the experimental values. From Table~\ref{lambda_lec}, it is evident that only the leading-order term in the chiral expression is determined well.

\begin{center}
\begin{table}
\begin{tabular}{|cccccccccccc|ccc|}
\hline
\hline
${m^0_{\Lambda_c}}/{m_\Omega^0}$& ${\Delta^0_{\Lambda_c\Sigma_c}}/{m_\Omega^0}$&${\Delta^0_{\Lambda_c\Sigma^*_c}}/{m_\Omega^0}$&$\bar\sigma_{\Lambda_c}$&$\bar\sigma_{\Sigma_c}$&$\bar\sigma_{\Sigma^*_c}$&$g_{3}^2$&$g_{2}^2$&$c_a/{m_\Omega^{\rm{phys}}}$&$\chi^2$&d.o.f.&$Q$\\
\hline
1.352(33)&  0.112(30)&   0.162(72)   &   1.3(1.7)& 1.2(5.2)&1(15)&  0.2(4.9)& 0(16)&  0.0042(71)
&6.4
&6&0.4
\\
\hline
\hline
\end{tabular}
\caption{Results of $SU(2)$ HH$\chi$PT LECs from fits of the $\{\Lambda_c,\Sigma_c,\Sigma^*_c\}$ multiplet masses, $\chi^2$, the number of degrees of freedom, and the goodness of the fit ${Q(d)}$ (as defined in Sec.~\ref{fitting}).
}
\label{lambda_lec}
\end{table}
\end{center}

\begin{figure}
\begin{center}
\includegraphics[totalheight=10cm]{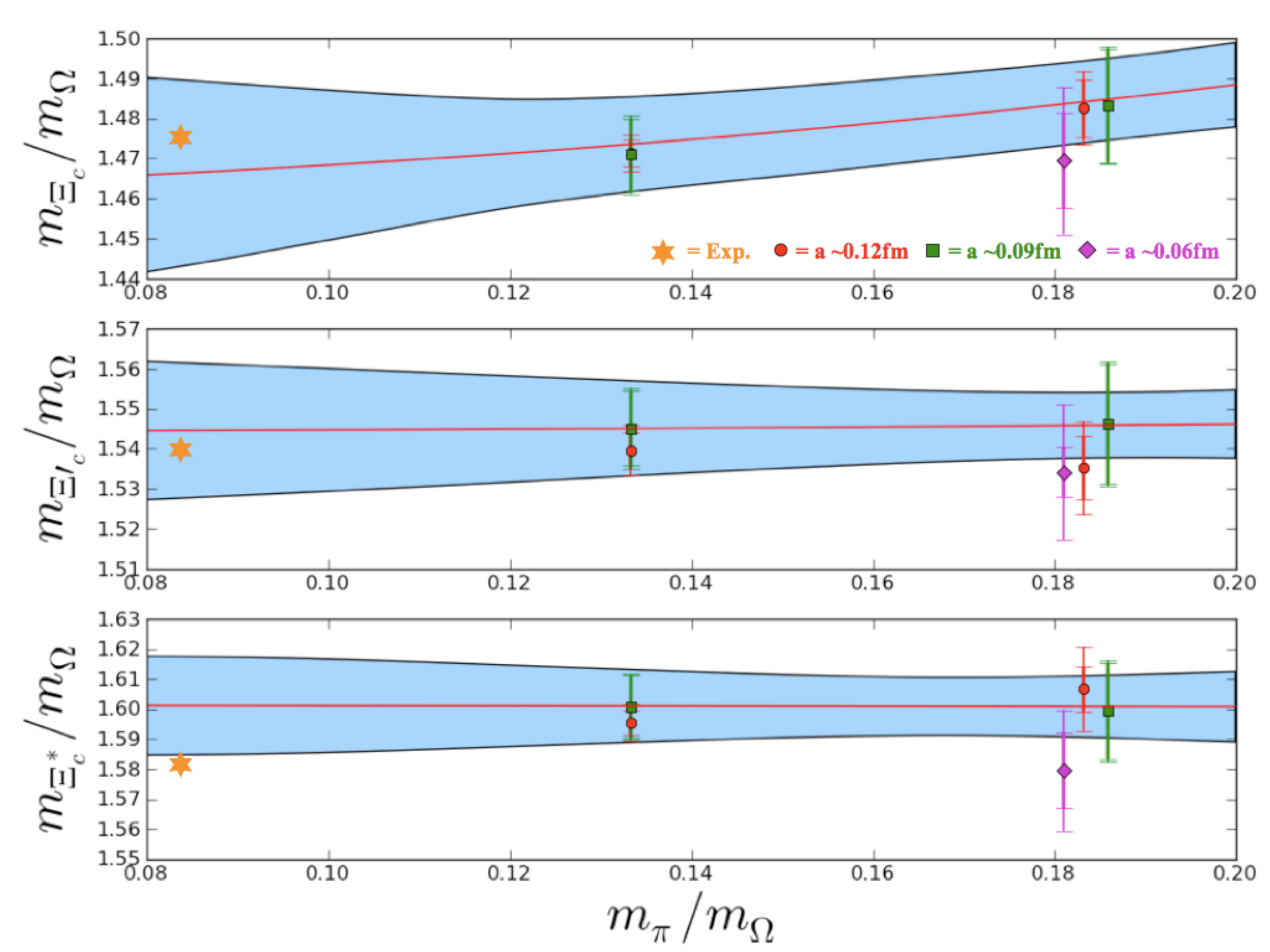}
\end{center}
\caption{NLO HH$\chi$PT and continuum extrapolations of $\{\Xi_{c},\Xi'_{c},\Xi^*_{c}\}$  masses. The red line depicts the fit of the data that has been extrapolated to $a=0$. The blue band includes the statistical and systematic errors added in quadrature. }\label{chic_plot}
\end{figure}

Next consider the multiplet $\{\Xi_c,\Xi'_c,\Xi^*_c\}$:
\begin{eqnarray}
\label{xicc}
\frac{m_{\Xi_c}}{m_\Omega}&=&\frac{m^0_{\Xi_c}}{m_\Omega^0}+\frac{\bar\sigma_{\Xi_c}m^2_{\pi}}{(4\pi f_\pi)m_\Omega}
-\frac{3}{2}\frac{g_3^2}{{(4\pi f_\pi)^2m_\Omega}}\left(\frac{1}{3}\mathcal{F}(m_\pi,\Delta_{\Xi_c\Xi'_c},\mu)+
\frac{2}{3}\mathcal{F}(m_\pi,\Delta_{\Xi_c\Xi^*_c},\mu)\right)\\
\frac{m_{\Xi'_c}}{m_\Omega}&=&\frac{m^0_{\Xi_c}+\Delta_{\Xi_c\Xi'_c}}{m_\Omega^0}+\frac{\bar\sigma_{\Xi'_c}m^2_{\pi}}{(4\pi f_\pi)m_\Omega}
-\frac{1}{2}\frac{g_3^2}{{(4\pi f_\pi)^2m_\Omega}}\mathcal{F}(m_\pi,-\Delta_{\Xi_c\Xi'_c},\mu)
+\frac{3}{8}\frac{g_2^2}{{(4\pi f_\pi)^2m_\Omega}}\left(\frac{4}{9}\mathcal{F}(m_\pi,0,\mu)+
\frac{8}{9}\mathcal{F}(m_\pi,\Delta_{\Xi'_c\Xi^*_c},\mu)\right)\nn\\
\frac{m_{\Xi^*_c}}{m_\Omega}&=&\frac{m^0_{\Xi_c}+\Delta_{\Xi_c\Xi^*_c}}{m_\Omega^0}+\frac{\sigma_{\Xi^*_c}m^2_{\pi}}{(4\pi f_\pi)m_\Omega}
-\frac{1}{2}\frac{g_3^2}{{(4\pi f_\pi)^2m_\Omega}}\mathcal{F}(m_\pi,-\Delta_{\Xi_c\Xi^*_c},\mu)
+\frac{3}{8}\frac{g_2^2}{{(4\pi f_\pi)^2m_\Omega}}\left(\frac{10}{9}\mathcal{F}(m_\pi,0,\mu)+
\frac{2}{9}\mathcal{F}(m_\pi,-\Delta_{\Xi'_c\Xi^*_c},\mu)\right)
.\nn
\end{eqnarray}
The values obtained for the LECs are shown in Table~\ref{chic_lec}, and Fig.~\ref{chic_plot} displays our fits at the continuum. From Fig.~\ref{chic_plot} it is clear that the extrapolated masses are within $1.1~\sigma$ from the experimental values. Furthermore, we are not able to resolve any lattice-spacing dependence for this multiplet, and the chiral extrapolation is close to a constant for $\Xi'_c$ and $\Xi^*_c$.

\begin{center}
\begin{table}
\begin{tabular}{|cccccccccccc|ccc|}
\hline
\hline
${m^0_{\Xi_{c}}}/{m_\Omega^0}$& ${\Delta^0_{\Xi_{c}\Xi'_{c}}}/{m_\Omega^0}$&${\Delta^0_{\Xi_{c}\Xi^*_{c}}}/{m_\Omega^0}$&$\bar\sigma_{\Xi_{c}}$&$\bar \sigma_{\Xi'_{c}}$&$\bar \sigma_{\Xi^*_{c}}$&$g_{3}^2$&$g_{2}^2$&$c_a/{m_\Omega^{\rm{phys}}}$&$\chi^2$&d.o.f.&$Q$\\
\hline
1.477(45)&0.054(63)& 0.11(16)&0.73(60)& 0.1(6.7)&$-$0.4(5.1)&  3.0(7.1)&   0.0(6.4)&  0.006(10)

&5.2&6&0.5\\

\hline
\hline
\end{tabular}
\caption{Results of $SU(2)$ HH$\chi$PT LECs from fits of the $\{\Xi_{c},\Xi'_{c},\Xi^*_{c}\}$ masses. }
\label{chic_lec}
\end{table}
\end{center}

\begin{figure}
\begin{center}
\includegraphics[totalheight=10cm]{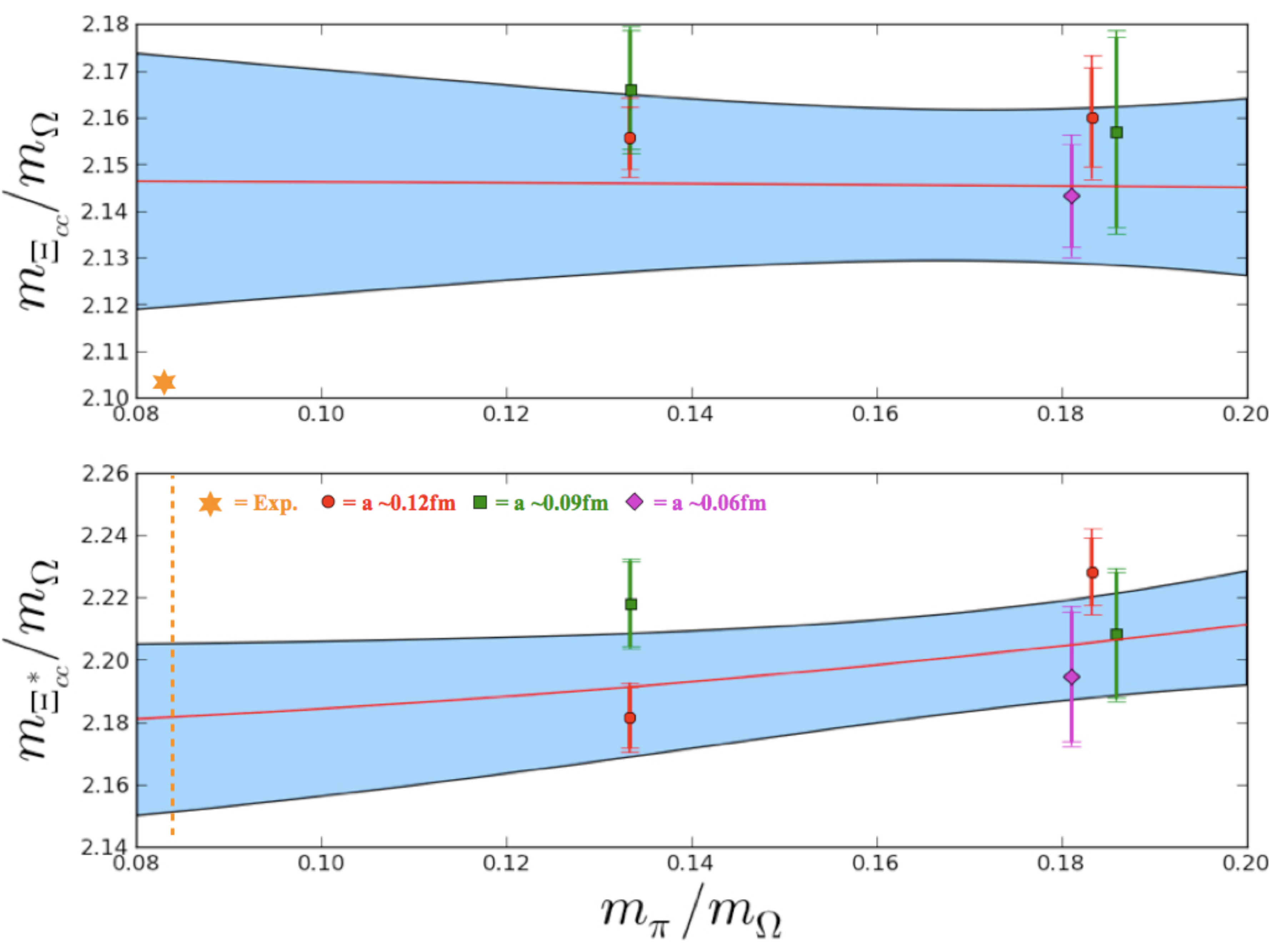}
\end{center}
\caption{NLO HH$\chi$PT and continuum extrapolations of $\{\Xi_{cc}, \Xi^*_{cc}\}$  masses. The red line depicts the fit of the data that has been extrapolated to $a=0$. The blue band includes the statistical and systematic errors added in quadrature. The dashed line indicates the physical point $m_\pi/m_\Omega=0.083453(25)$. }\label{chicc_plot}
\end{figure}

For the multiplet $\{\Xi_{cc}, \Xi^*_{cc}\}$ we use the previously determined expressions~\cite{su22} to perform the chiral extrapolation
\begin{eqnarray}
\frac{m_{\Xi_{cc}}}{m_\Omega}&=&\frac{m^0_{\Xi_{cc}}}{m_\Omega^0}
+\frac{\bar\sigma_{\Xi_{cc}}m^2_{\pi}}{(4\pi f_\pi)m_\Omega}
-\frac{g_{\pi\Xi_{cc} \Xi^*_{cc}}^2}{{(4\pi f_\pi)^2}{m_\Omega}}\left[\frac{1}{9}\mathcal{F}(m_\pi,0,\mu)+\frac{8}{9}\mathcal{F}(m_\pi,\Delta_{\Xi_{cc} \Xi^*_{cc}},\mu)\right]+ \mathcal{O}(m_\pi^4)\\
\frac{m_{\Xi^*_{cc}}}{m_\Omega}&=&\frac{m^0_{\Xi_{cc}}+\Delta_{\Xi_{cc} \Xi^*_{cc}}}{{m_\Omega^0}}
+\frac{\bar\sigma_{ \Xi^*_{cc}}}{(4\pi f_\pi)}{m^2_{\pi}m_\Omega}
-\frac{g_{\pi\Xi_{cc} \Xi^*_{cc}}^2}{{(4\pi f_\pi)^2}{m_\Omega}}\left[\frac{5}{9}\mathcal{F}(m_\pi,0,\mu)+\frac{4}{9}\mathcal{F}(m_\pi,-\Delta_{\Xi_{cc} \Xi^*_{cc}},\mu)\right]+ \mathcal{O}(m_\pi^4)\nn.
\end{eqnarray}
The results for the LECs are shown in Table~\ref{chicc_lec}, and Fig.~\ref{chicc_plot} displays our fits at the continuum. It is remarkable in Fig.~\ref{chicc_plot} that the $m_\pi^2$-dependence of $m_{\Xi_{cc}}$ is surprisingly small compared to that of $m_{\Xi^*_{cc}}$. From Fig.~\ref{chicc_plot}, one can also observe that our value of $m_{\Xi_{cc}}$ is about $1.7~\sigma$ above the experimentally observed mass.

\begin{center}
\begin{table}
\begin{tabular}{|cccccc|ccc|}
\hline
\hline
${m^0_{\Xi_{cc}}}/{m_\Omega^0}$& ${\Delta^0_{\Xi_{cc}\Xi_{cc}^*}}/{m_\Omega^0}$&$\bar\sigma_{\Xi_{cc}}$&$\bar \sigma_{\Xi_{cc}^*}$&$g_{\pi\Xi_{cc}\Xi_{cc}^*}^2$&$c_a/{m_\Omega^{\rm{phys}}}$&$\chi^2$&d.o.f.&$Q$\\
\hline
2.147(35)&0.025(24)&$-$0.00002(55)&0.00057(60)&0.00008(52)&0.013(19)& 6.3&4&0.2

\\
\hline
\hline
\end{tabular}
\caption{Results of $SU(2)$ HH$\chi$PT LECs from fits of the $\{\Xi_{cc},\Xi^*_{cc}\}$ multiplet masses. }
\label{chicc_lec}
\end{table}
\end{center}

\begin{figure}
\begin{center}
\includegraphics[totalheight=10cm]{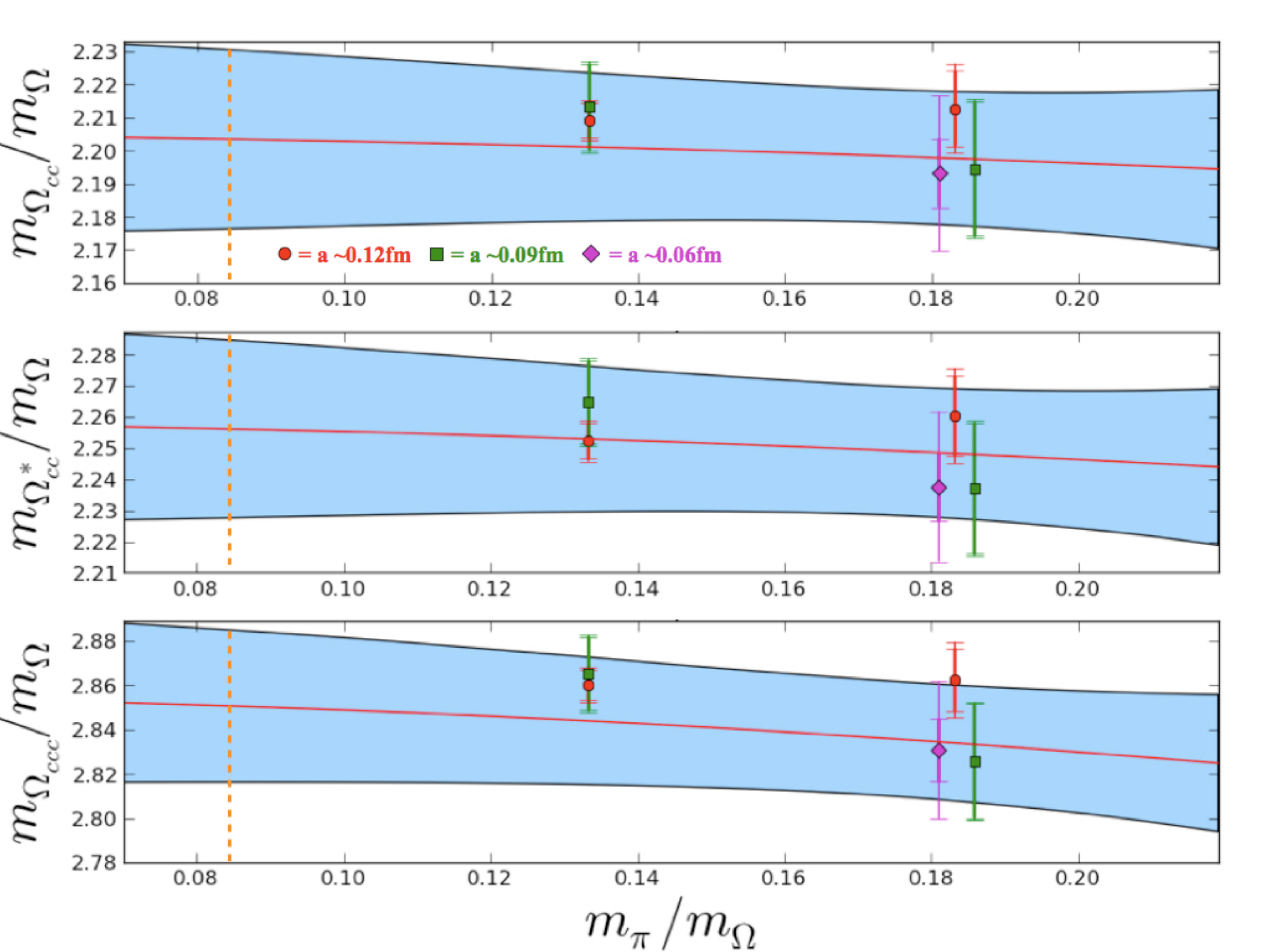}
\end{center}
\caption{Chiral and continuum extrapolations of $\{\Omega_{cc}, \Omega_{cc}^*, \Omega_{ccc}\}$ masses. The red line depicts the fit of the data that has been extrapolated to $a=0$. The blue band includes the statistical and systematic errors added in quadrature. The dashed line indicates the physical point $m_\pi/m_\Omega=0.083453(25)$. }\label{omegas}
\end{figure}

Lastly, the $SU(2)$ HH$\chi$PT extrapolation formula for all isosinglet states, $\Omega_c,\Omega^*_c, \Omega_{cc},\Omega^*_{cc},$ and $\Omega_{ccc}$, is given by
\begin{eqnarray}
\frac{m_{\Omega_{c}}}{m_\Omega}&=&\frac{m^0_{\Omega_{c}}}{m_\Omega^0}
+\frac{\bar\sigma_{{\Omega_{c}}}m^2_{\pi}}{(4\pi f_\pi)m_\Omega} + \mathcal{O}(m_\pi^4).
 \end{eqnarray}
In Table~\ref{omega_lec}, we summarize the fitted LECs of the five isosinglet states. Figure~\ref{omegas} shows the continuum extrapolation of the yet-to-be-observed $\{\Omega_{cc},\Omega_{cc}^*,\Omega_{ccc}\}$ states along with the value of the ratio of their masses to $m_\Omega$ for each ensemble.

\begin{center}
\begin{table}
\begin{tabular}{|c|ccc|ccc|}
\hline
\hline
Hadron&$m^0_{H}/{m_\Omega^0}$&$\bar\sigma_{H}$&$c_a/{m_\Omega^{\rm{phys}}}$&$\chi^2$&$\text{d.o.f}$&$Q$\\
\hline
$\Omega_c$&
1.612(24)&
$-$0.49(66)&
$-$0.005(18)&0.57&2&0.57\\
$\Omega^*_{c}$&
1.670(23)&
$-$0.78(62)&
$-$0.005(18)&1.32&2&0.27\\
$\Omega_{cc}$&
2.206(30)&
$-$0.27(81)&
0.010(24)&0.58& 2&0.56\\
$\Omega^*_{cc}$&
2.247(33)&
$-$0.17(88)&
0.010(26)&0.81&2&0.44
\\
$\Omega_{ccc}$&
2.857(38)&
$-$0.7(1.0)&
0.019(29)&1.16
&2&0.31
\\
\hline
\hline
\end{tabular}

\caption{LO $SU(2)$ $\chi$PT LECs of isosinglet states $\Omega_c,\Omega^*_c, \Omega_{cc},\Omega^*_{cc},$ and $\Omega_{ccc}$.}
\label{omega_lec}
\end{table}
\end{center}

\subsection{Systematics \label{system}}

In performing the continuum and chiral extrapolation, we added five systematic errors in addition to the fitting-window error. The first of these arises from the uncertainty in determining $m_\pi$ and the lattice spacing. We derive this uncertainty by simultaneously varying $m_\pi$ and the lattice spacing within their corresponding uncertainties (shown in Tables~\ref{ensembles} and~\ref{ensembles2}, respectively) when extrapolating the masses to the physical point. This gives an ensemble of energies, and we obtain a systematic uncertainty from the standard deviation of this ensemble.

The second uncertainty is due to finite-volume (FV) corrections. The dominant finite-volume effects for baryon with light degrees of freedom from the FV counterpart of self-energy diagrams depicted in Fig.~\ref{se}, and in the p-regime these scale like $e^{-m_\pi L}/(m_\pi L)$~\cite{pregime}. More specifically, up to an overall $\mathcal{O}(1)$ constant, they can be written as~\cite{pregime}
\begin{eqnarray}
\delta m_H^{{\rm FV},l}\sim \frac{m_\pi^3}{8\pi f^2_\pi}\sum_{\vec{n}\neq\vec{0}}\frac{e^{-L{|\vec{n}|}{m_\pi}}}{m_\pi L{|\vec{n}|}}.
\end{eqnarray}
Note, the overall constant depends on the axial coupling, which we have found to be consistent with zero (see Tables~\ref{lambda_lec},~\ref{chic_lec} and~\ref{chicc_lec}). For hadrons with no light degrees of freedom, FV effects come in at $\mathcal{O}(m_\pi^4)$ in the chiral expansion, and therefore are further suppressed by a factor of $m_\pi/\Lambda_{\chi}$, where $\Lambda_{\chi}\sim 700$~MeV is the chiral symmetry-breaking scale,
\begin{eqnarray}
\delta m_H^{{\rm FV},h}\sim \frac{m_\pi^4}{8\pi f^2_\pi\Lambda_{\chi}}\sum_{\vec{n}\neq\vec{0}}\frac{e^{-L{|\vec{n}|}{m_\pi}}}{m_\pi L{|\vec{n}|}}.
\end{eqnarray}
In Table~\ref{systematics} we evaluate both of these FV effects for hadrons with and without light degrees of freedom.

In performing the chiral and continuum extrapolation we have taken into account terms coming in at $\mathcal{O}(a^2, m_\pi^2, m_\pi^3,1/m_Q)$ and neglected $\mathcal{O}( m_\pi^4,a^2m_\pi)$ terms. In order to account for $\mathcal{O}( m_\pi^4)$ corrections we add a systematics of the form~\cite{hhchipt7}
\begin{eqnarray}
\delta m_H^{\chi \text{PT}}\sim \frac{m^4_\pi}{(4\pi f_\pi)^3},
\end{eqnarray}
which contributes at the MeV level for our ensembles.

In general, quantities obtained using mixed action have discretization errors arising from artifacts of both the sea and the valence actions. From mixed-action EFT (MAEFT) we know that leading order (LO) these artifacts can be parametrized in terms of two quantities, $a^2\Delta_{\rm Mix}$ and  $a^2\Delta_{\rm sea}$~\cite{ Bar:2005tu, Bar:2002nr}. The latter contributes to the LO $a$-dependence of the valence pion mass, which have been accounted for in our continuum extrapolation. Both of these splittings give rise to NLO corrections to the MAEFT extrapolation formulas. In Ref.~\cite{Orginos:2007tw}, Orginos and Walker-Loud evaluated $a^2\Delta_{\rm Mix}$ for the asqtad improved MILC lattices with $a\approx0.125$~fm and found it to be $(316(4)\mbox{ MeV})^2$, which is smaller than the corresponding value of $a^2\Delta_{\rm sea}=(450\mbox{ MeV})^2$. To this day, $a^2\Delta_{\rm Mix}$ has not been determined for the HISQ MILC lattices. Assuming this hierarchy ($a^2\Delta_{\rm sea}\geq a^2\Delta_{\rm Mix}$) is also true for the HISQ lattices, we can use power-counting arguments to estimate the $\mathcal{O}(a^2m_\pi)$ corrections,
\begin{eqnarray}
\delta m_H^{\rm MA}\sim  m_\pi \frac{a^2\Delta_{\rm sea}}{(4\pi f_\pi)^2}.
\end{eqnarray}
The values of the $a^2\Delta_{\rm sea}$ splittings, which is the mass difference between the Goldstone Kogut-Susskind sea pion and the staggered taste-singlet meson, have been determined numerically by the MILC Collaboration for the ensembles we are using~\cite{MILC, MILC2}. From these values we obtain the $\delta m_H^{\rm MA}$ shown in Table~\ref{systematics}. Note, the overall $\mathcal{O}(1)$ constants present in this correction depend on the axial coupling.

Furthermore, since we have used the strange mass to set the scale, we need to account for possible mismatch between the sea and valence strange-quark masses. We use power-counting arguments to estimate the leading-order correction:
\begin{eqnarray}
\delta m_H^{s}\sim   \frac{{\left|(m^2_K)_{\rm{val}}-(m^2_K)_{\rm{sea}}\right|}}{(4\pi f_K)},
\end{eqnarray}
where $f_K=156.1(9)$~MeV is the kaon decay constant. One can certainly include a similar error for the light-quark mismatch, but this would be below our level of precision ($0.1$~MeV).

We then add these five sources of systematics for each ensemble and extrapolate them to the physical point, which is shown as the third uncertainty of the physical masses in Table~\ref{results_all}.

\begin{center}
\begin{table}
\begin{tabular}{|c|ccccc|}
\hline

& $\hspace{.1cm}\delta m_H^{{\rm FV},l}[\text{MeV}]\hspace{.1cm}$& $\hspace{.1cm}\delta m_H^{{\rm FV},h}[\text{MeV}]\hspace{.1cm}$&$\hspace{.1cm}\delta m_H^{\chi\text{PT}}[\text{MeV}]\hspace{.1cm}$&$\hspace{.1cm}\delta m_H^{{\rm MA}}[\text{MeV}]\hspace{.1cm}$&$\hspace{.1cm}\delta m_H^{s}[\text{MeV}]\hspace{.1cm}$\\
\hline \hline
$\textbf{A1}$& 1.3&0.6& 2.1& 4.5 &0.2\\\hline
${\textbf{A2}}$ & 0.5&0.2& 1.0& 3.2&0.7 \\\hline
${\textbf{B1}} $& 1.3&0.6& 2.3&2.8&0.6 \\\hline
${\textbf{B2}} $& 0.3& 0.1&0.6& 2.0&2.3\\\hline
${\textbf{C1}} $& 1.3&0.6& 2.1&0.5&2.1\\\hline\hline
\end{tabular}
\caption{Shown are estimates for the systematic errors for each ensemble. From left to right column, they are the systematic errors due to finite-volume effects for baryons with light degrees of freedom, finite-volume effects for hadrons with no light degrees of freedom, the truncation of the $\chi$PT extrapolation formulas, corrections in the MAEFT expansion, and the sea/valence strange-mass mismatch, respectively. }
\label{systematics}
\end{table}
\end{center}

 \section{Discussion and Conclusion \label{conclusion}}

\begin{figure}
\begin{center}
\includegraphics[totalheight=12cm]{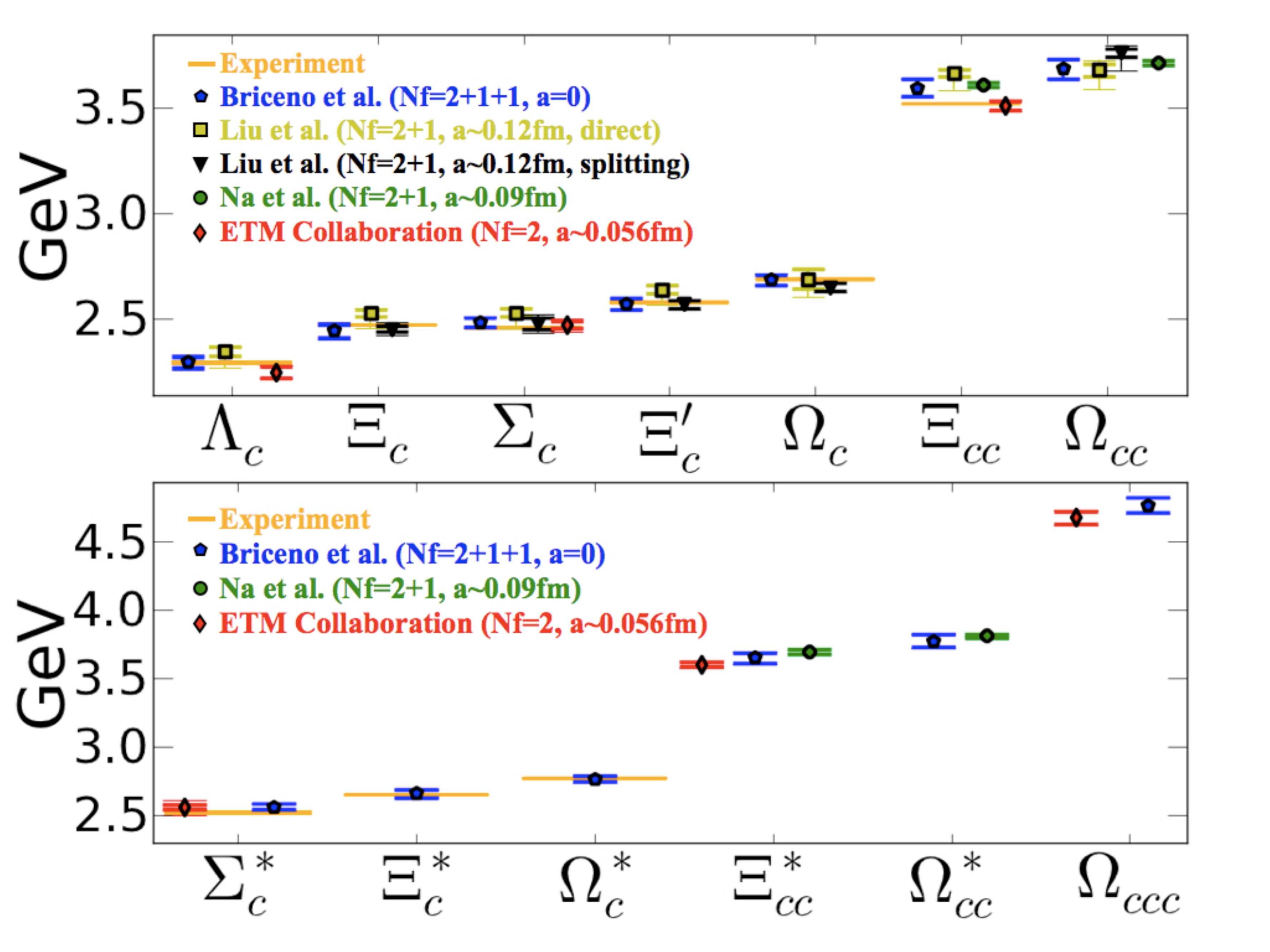}
\end{center}
\caption{A survey of previous unquenched lattice calculations~\cite{latt1, latt15, latt16, latt2, latt25, Alexandrou:2012xk}, along with the results of this paper labeled as ``Briceno~et~al.'' The ``$a=0$'' label denotes calculations that extrapolated their results to the continuum limit. The statistical uncertainty is shown as a thick inner error bar, while the statistical and systematic uncertainties added in quadrature are shown as a larger thin outer error bar. Our systematic uncertainties include errors originating from the fitting window and scale setting. The experimentally determined masses are shown for comparison~\cite{pdg}. }\label{Spectrum}
\end{figure}
\begin{center}
\begin{table}
\begin{tabular}{c|cc||c|cc}
\hline
\hline

Hadron& Latt. Pred.~[MeV]& Exp.~[MeV]&Hadron& Latt. Pred. [MeV]& Exp. [MeV]\\
$\eta_c$&$2995(26)(12)(5)$
&2980.3(1.2)&
$\Sigma_c$&2481(24)(15)(7)&2454.02(2)\\
$J/\psi$&$3092(27)(13)(6)$
&3096.916(11)&
$\Sigma^*_c$&2559(30)(15)(7)&2518.4(6)\\
$\chi_{c0}$&$3397(31)(15)(6)$&3414.75(31)&
$\Xi'_c$&2568(25)(12)(6)
&2575.6(3.1)\\
$\chi_{c1}$&$3540(38)(19)(5)$
&3510.66(7)&
$\Xi^*_c$&2655(26)(6)(7)&2645.9(6)\\
$h_c$&$3559(37)(18)(6)$&3525.41(16)&
$\Omega_c$&2681(31)(12)(5)&2685.2(1.7)\\
$\Delta_{1S}$&$110.9(1.1)(1.4)(5.3)$&116.6(1.2)&
$\Omega^*_c$&2764(30)(14)(5)&2765.9(2.0)\\
$D_s$&1960(17)(18)(5)&1968.45(33)&
$\Xi_{cc}$&3595(39)(20)(7)
&3518.9(9)\\
${D_s-\eta_c/2}$&468.7(4.8)(5.6)(5.8)&478.30(69)&
$\Xi^*_{cc}$&3648(42)(18)(7)
&---\\
$K^+$&488.7(5.3)(5.3)(5.8)&493.677(16)&
$\Omega_{cc}$&3679(40)(17)(5)&---\\
$\Lambda_c$&2291(37)(22)(7)&2286.46(14)&
$\Omega^*_{cc}$&3765(43)(17)(5)&---\\
$\Xi_c$&2439(29)(25)(7)
&2467.8(6)&
$\Omega_{ccc}$&4761(52)(21)(6)&---\\

\hline
 \end{tabular}
\caption{Results for the charmed hadron spectrum after extrapolating the masses in Tables~\ref{cc_ens} and~\ref{baryon_ens} to the physical point. The first uncertainty is due to statistics, the second to the fitting window error, and the third corresponds to scale setting, finite-volume effects, $\mathcal{O}(m^4_\pi,a^2m_\pi)$-corrections to the expressions used to extrapolate to the physical point, and strange-mass tuning  errors added in quadrature (as discussed in Sec.~\ref{system}).}

\label{results_all}
\end{table}
\end{center}
\begin{figure}
\begin{center}
\includegraphics[totalheight=8cm]{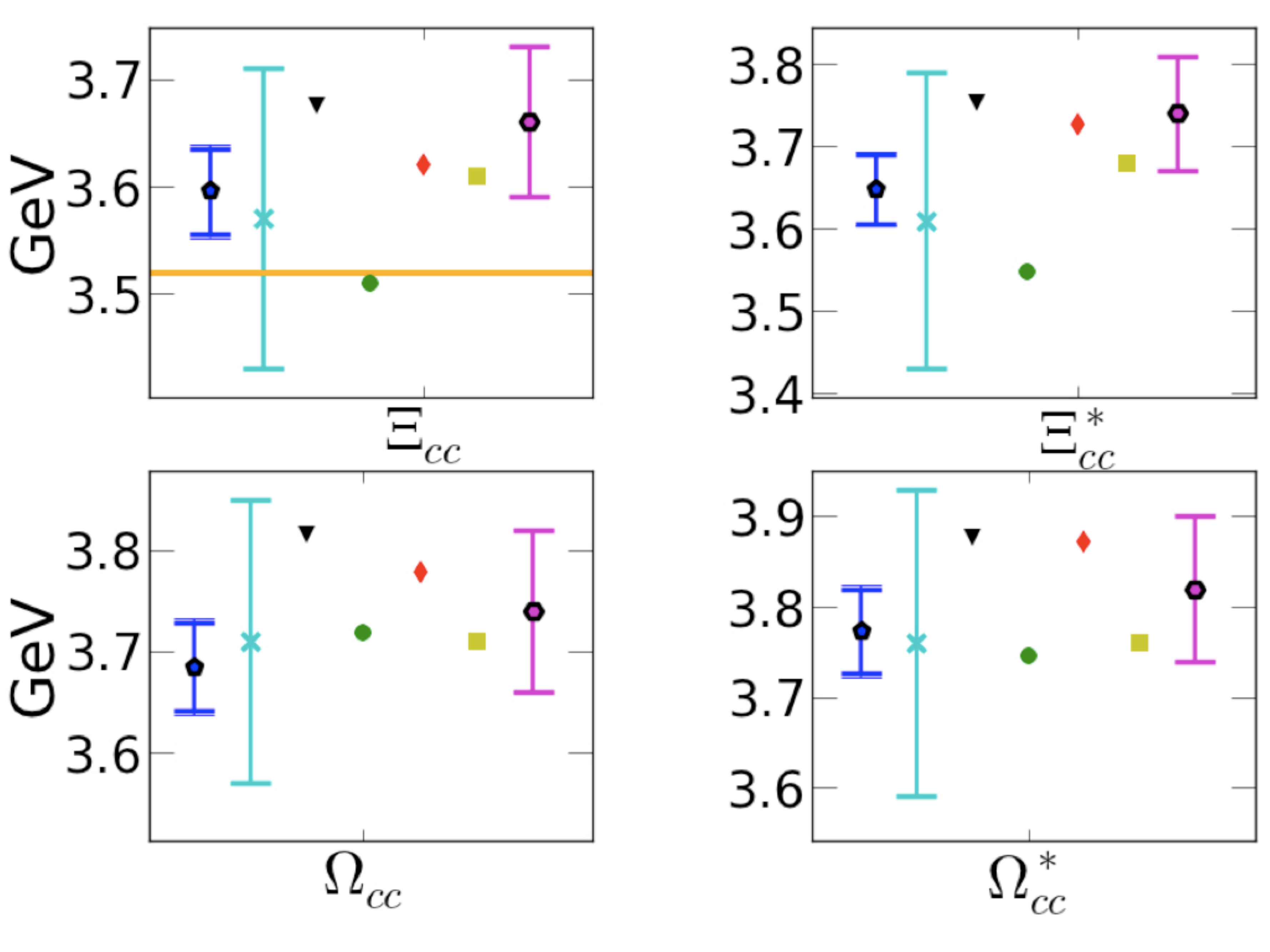}
\includegraphics[totalheight=8cm]{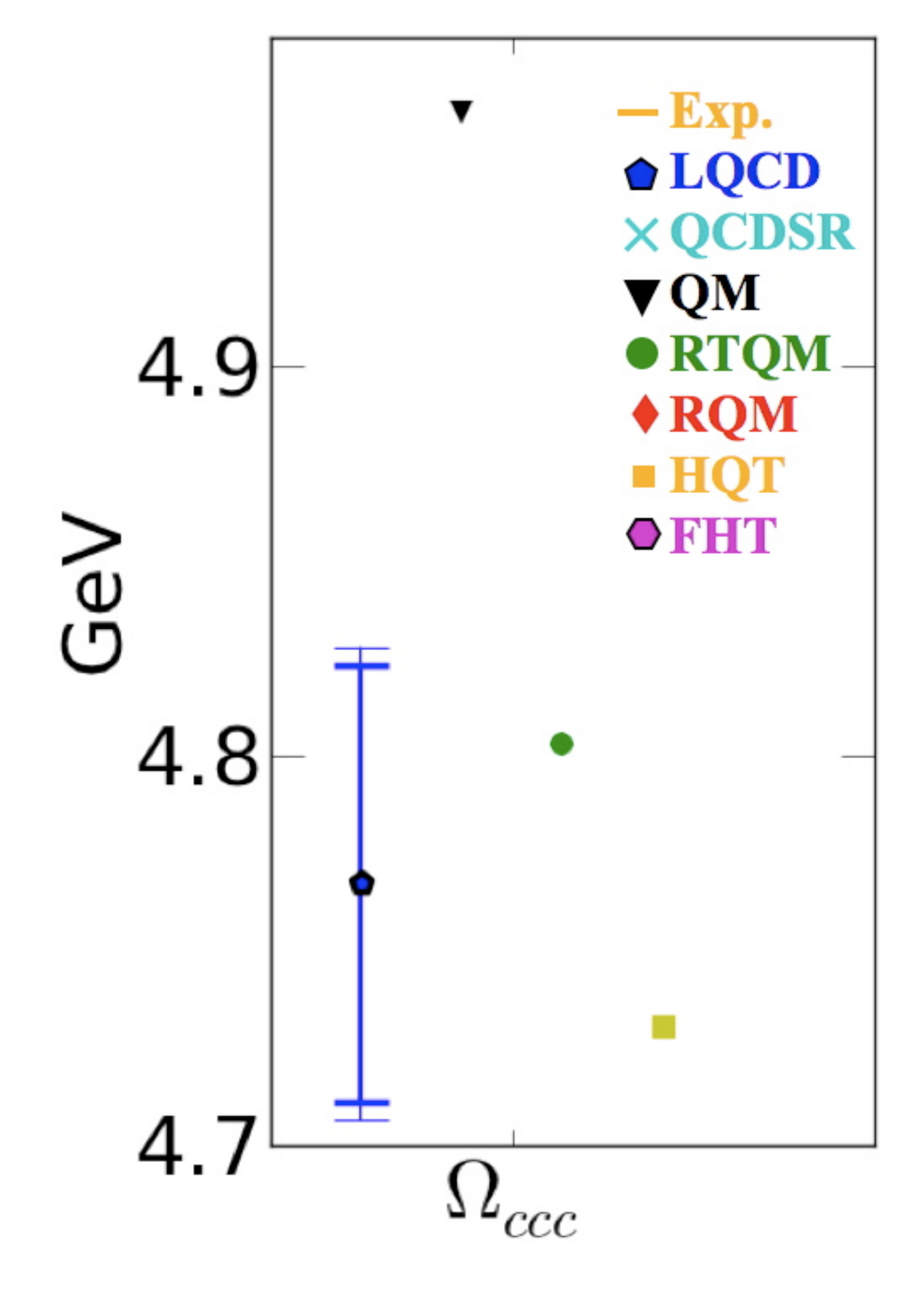}
 \end{center}
\caption{Comparison of our results (LQCD) for the masses of the lightest doubly and triply charmed baryons, with the theoretical prediction from other models: QCD sum rules (QCDSR)~\cite{QCDsum, QCDsum2}, the nonrelativistic quark model (QM)~\cite{QM}, the relativistic three-quark model (RTQM)~\cite{RTQM}, the relativistic quark model (RQM)~\cite{RQM}, heavy-quark effective theory (HQET)~\cite{HQET}, and the Feynman-Hellmann theorem (FHT)~\cite{FHT}. }
\label{modelsf}
\end{figure}

In this work we presented the first unquenched continuum determination of the low-lying charmed-baryon spectrum. The calculation uses a relativistic heavy-quark action for the valence charm quark, clover-Wilson fermions for the valence light and strange quarks, and HISQ sea quarks generated by the MILC Collaboration ~\cite{MILC, MILC2}. The spectrum is calculated with a lightest pion mass around $220$~MeV, and three lattice spacings ($a\approx0.12$~fm, $0.09$~fm, and $0.06$~fm) are used to extrapolate to the continuum. At each ensemble, we interpolate the charm-quark mass to the physical one by matching the charmonium $1S$ spin average through the ratio
$({m^{\rm{phys}}_{\eta_c}+3m^{\rm{phys}}_{J/\psi}})/{(4m^{\rm{phys}}_{\Omega})}=1.83429(56)$; the rest of the hadron (composed of charm quarks) ratios $m_H/m_{\Omega}$  are
linearly interpolated in $am_{c}$ to the physical charm point.

In order to determine the lattice spacing for the five ensembles, we chose to use the $\Omega$ mass due to its weak $m_\pi$-dependence. This was done by extrapolating the value of $am_\Omega$ over all ensembles with the same value of $\beta$ to the physical pion mass. We then obtained the lattice spacing by dividing $am_\Omega$ by the physical $\Omega$ mass. The resulting values of the lattice spacing are shown in Table~\ref{ensembles2}.

The main result of this work is the charmed hadron spectrum shown in Table~\ref{results_all}, which was obtained by extrapolating measurements from the five ensembles to the physical point defined by $m^{\rm{phys}}_\pi/m^{\rm{phys}}_\Omega=0.083453(25)$ and $a=0$~\cite{pdg}. When performing the chiral and continuum extrapolation we use HH$\chi$PT up to $\mathcal{O}(m^3_\pi,1/m_c,a^2)$. The three uncertainties of the masses shown correspond to statistics, fitting-window error, and systematics from other lattice artifacts, such as lattice-scale setting and pion-mass determination (as discussed in Sec.~\ref{system}).

To test our tuning and extrapolation procedure, we verify that our calculation does in fact reproduce the well known low-lying $ l\bar{s}, c\bar{s},c\bar{c}$ spectrum. Since we use the strange quark mass to set the scale, we first determine the kaon mass. As is shown in Figs.~\ref{kaon_plot}, after extrapolating to the physical point we obtain $m_{K^+}=488.7(5.3)(5.3)(5.7)$~MeV, which is in perfect agreement with experiment and it displays a minimal lattice-spacing dependence. The remaining results for the $c\bar{s},c\bar{c}$ spectrum are shown in Figs.~\ref{charmonium} and~\ref{deltasc}, and it is evident that we do in fact recover the physical spectrum in the mesonic sector. Two particularly interesting quantities are the $D_s$-$\eta_c/2$ splitting and the $\Delta_{1S}$, both of which show significant $a$-dependence. The fact that we only obtain agreement with experiment after extrapolating to the continuum confirms the necessity of performing calculations of the charmed spectrum at multiple lattice spacings.

In Fig.~\ref{Spectrum} we display the results for the charmed-baryon spectrum, along with a survey of previous unquenched lattice calculations~\cite{latt1, latt15, latt16, latt2, latt25, Alexandrou:2012xk} and corresponding experimental values for comparison~\cite{pdg}. Liu~et~al.~\cite{latt1, latt15, latt16} evaluated the charmed-baryon spectrum for four different pion masses (with lowest $m_\pi\approx 290$~MeV) and a single lattice spacing $a\approx0.125$~fm. Na~et~al.~\cite{latt2, latt25} evaluated the charmed-baryon spectrum at three different lattice spacings ($a\approx0.15$~fm, $0.12$~fm, and $0.09$~fm) but have yet to present extrapolated values of the masses as well as an estimate of their systematic uncertainties. The European Twisted Mass (ETM) Collaboration recently presented determined the masses of $\Lambda_c$, $\Sigma_c$, $\Sigma^*_c$, $\Xi_{cc}$, $\Xi^*_{cc}$, and $\Omega_{ccc}$,
using $N_f=2$ dynamical sea quarks with a lightest pion mass of $260$~MeV and three lattice spacings $a\in \{0.056(1),0.0666(6), 0.0885(6)\}$~fm~\cite{Alexandrou:2012xk}. Despite having performed calculations at three different lattice spacings, they did not extrapolate results to the continuum. Instead, they used the information gathered by using multiple lattice spacing to quantify the discretization error of their calculations. The use of $N_f=2$ dynamical quarks, introduces a source of systematic error that is hard to quantify and has not been addressed by the ETM Collaboration. That being said, Fig.~\ref{Spectrum} shows that the masses calculated by the ETM Collaboration are in agreement with our results with the exception of the controversial $\Xi_{cc}$, where our result is about $1.6~\sigma$ above the value obtained by the ETM Collaboration.

All previous calculations of the charmed-baryon spectrum have been performed with light-quark masses corresponding to $m_\pi\geq 260$~MeV, placing our calculation closest to the physical point. Furthermore, ours is the only  determination of the continuum charmed-baryon spectrum. Perhaps the most pertinent of the results presented is the $\Xi_{cc}$ mass, 3595(39)(20)(6)~MeV. Unlike all previous calculations, we performed a coupled extrapolation of the $\{\Xi_{cc},\Xi_{cc}^*\}$ doublet to the physical point. Although, this led to a $m_{\Xi_{cc}}$ closer to the experimentally observed value in comparison to our previous work~\cite{Briceno}, our mean value of $m_{\Xi_{cc}}$ is still above the mass observed by the SELEX Collaboration by about $76$~MeV~\cite{SELEX1, SELEX2} and our combined uncertainty for this particle is $44$~MeV. Therefore, despite the fact that we see no strong disagreement with the SELEX result, our result does not agree with their experimentally observed mass. This is in contrast with the recently published result by the ETM Collaboration, $m_{\Xi_{cc}}=3513(23)(14)$~\text{MeV}~\cite{Alexandrou:2012xk}, which is the only unquenched LQCD calculation to be in agreement with the SELEX Collaboration.

Therefore, it remains true that the $\Xi^+_{cc}$ requires further investigation both from the experimental and the theoretical communities. In particular, from the experimental side it would be desirable to obtain a clear determination of the isospin doublet $(\Xi^+_{cc},\Xi^{++}_{cc})$ masses. Although, the SELEX Collaboration has confirmed their observation of $\Xi^+_{cc}(3520)$, the BABAR~\cite{BABAR1}, BELLE~\cite{BELLE1}, and FOCUS~\cite{FOCUS} experiments observed no evidence for either state of the doublet.
From the theoretical side, we expect to be able to perform calculations closer to or at the physical pion mass in the near future, thereby reducing the contribution from lattice artifacts. In Fig.~\ref{modelsf} we compare our results for the masses of doubly and triply charmed baryons with predictions from theoretical models. In particular, we show results obtained using QCD sum rules (QCDSR)~\cite{QCDsum, QCDsum2}, the nonrelativistic quark model (QM)~\cite{QM}, the relativistic three-quark model (RTQM)~\cite{RTQM}, the relativistic quark model (RQM)~\cite{RQM}, heavy-quark effective theory (HQET)~\cite{HQET}, and the Feynman-Hellmann theorem (FHT)~\cite{FHT}. Our result for $m_{\Xi_{cc}}$ is 3595(39)(20)(6)~MeV, and from Fig.~\ref{modelsf} we estimate the overall theoretical prediction for this mass to be 3550--3650~MeV. These figures can guide experimentalists on the quest for the doubly and triply charmed-baryon masses. Finally, we predict the yet-to-be-discovered doubly and triply charmed-baryon masses $\Xi_{cc}^*$, $\Omega_{cc}$, $\Omega_{cc}^*$, $\Omega_{ccc}$ to be $3648(42)(18)(7)$~MeV, $3679(40)(17)(5)$~MeV, $3765(43)(17)(5)$~MeV and $4761(52)(21)(6)$~MeV, respectively.


\subsection*{Acknowledgment}

We thank MILC Collaboration and PNDME Collaboration
for sharing their HISQ lattices and light clover propagators with us.
RB thanks Martin Savage  for fruitful discussions, and for his feedback on the first manuscript of this paper. 
These calculations were performed using the Chroma software
suite \cite{chroma} on Hyak clusters at the University of Washington
eScience Institute, using hardware awarded by NSF grant
PHY-09227700. This work was made possible with the use of advanced computational, storage, and networking infrastructure provided by the Hyak supercomputer system, supported in part by the University of Washington eScience Institute.
The authors were supported by the DOE grant DE-FG02-97ER4014.


\bibliography{bibi}

\end{document}